\documentclass[twoside]{article}
\usepackage{latexsym,amsmath,amssymb}
\usepackage{hyperref}
\usepackage{cite} 
\usepackage{sectsty}
  \sectionfont{\large} \subsectionfont{\normalsize}
\usepackage{graphicx}

\usepackage{theorem}
\usepackage[all]{xy}
\newtheorem{theorem}{Theorem}[section]
{\theorembodyfont\rmfamily
\newtheorem{lemma}{Lemma}[section]

}
\newenvironment{proof}{\begin{paragraph}
          {Proof}}{\end{paragraph}}

\renewenvironment{abstract}
 {\small\begin{quote}{\textbf{Abstract}}\,\,}{\end{quote}}
\newenvironment{keywords}
 {\small\begin{quote}{\textbf{Keywords}}\,\,}{\end{quote}}
\newenvironment{classification}
 {\small\begin{quote}{\textbf{2010 Mathematics Subject Classification}}\,\,}{\end{quote}}
\date{}

\numberwithin{equation}{section}
\title{\vspace{-9ex}
{\centering
 \textbf{\large The Binomial Tree Method and Explicit Difference Schemes for American Options with Time Dependent Coefficients
}}}

 \vspace{-4ex}

\author{\small\textsf{\bfseries 
$^{1}$ Hyong-chol O, $^{2}$Song-gon Jang, $^{2}$Il-Gwang Jon, $^{2}$Mun-Chol Kim, $^{2}$Gyong-Ryol Kim, $^{2}$Hak-Yong Kim}\\[-.5ex]
{\footnotesize  ${}^{1, 2}$ Faculty of Mathematics, \textbf{Kim Il Sung} University,}
{\footnotesize   Pyongyang , D P R Korea}\\[-.5ex]
{\footnotesize e-mail:  $^{1}$hc.o@ryongnamsan.edu.kp }}

\begin{document}

\maketitle
\thispagestyle{empty}

\vspace{-.6cm}

\begin{abstract}
Binomial tree methods (BTM) and explicit difference schemes (EDS) for the variational inequality model of American options with time dependent coefficients are studied. When volatility is time dependent, it is not reasonable to assume that the dynamics of the underlying asset's price forms a binomial tree if a partition of time interval with equal parts is used. A time interval partition method that allows binomial tree dynamics of the underlying asset's price is provided. Conditions under which the prices of American option by BTM and EDS have the monotonic property on time variable are found. Using convergence of EDS for variational inequality model of American options to viscosity solution the decreasing property of the price of American put options and increasing property of the optimal exercise boundary on time variable are proved. First, put options are considered. Then the linear homogeneity and call-put symmetry of the price functions in the BTM and the EDS for the variational inequality model of American options with time dependent coefficients are studied and using them call options are studied.
\end{abstract}



%
%

\section{Introduction}

\indent 
There are two kinds of numerical methods for option pricing; one is based on the probabilistic approach and another one is the finite difference method for PDE. 

The binomial tree method (BTM), first proposed by Cox, Ross and Rubinstein \cite{CRR}, is one of the probabilistic numerical methods for pricing options. Due to its simplicity and flexibility, it has become one of the most popular approaches to pricing options. \cite{ami, AK, BPV, jia, KOF, LL, luo}

It is well known that the BTM for European option in Black - Scholes diffusion model converges to the corresponding continuous time model of Black and Scholes (\cite{he}). In particular, Jiang \cite{jia} showed that the BTM for European option is equivalent to a special explicit finite difference scheme for Black-Scholes PDE and proved its convergence using PDE approach. 

Amin and Khanna \cite{AK} first proved the convergence of BTM for American options using probabilistic approach. 

Jiang and Dai (\cite{JD1, JD2}) proved the convergence of explicit difference scheme and BTM for American options using viscosity solution theory of PDE. They showed that the BTM for American option is equivalent to a special explicit finite difference scheme for a variational inequality related to Black-Scholes PDE, proved monotonic property of the price by BTM and explicit finite difference scheme, existence and monotones of approximated optimal exercise boundary and used the method of Barles et al \cite{BDR, BS} and comparison principle of \cite{BDR, CIL}.  Jiang and Dai \cite{JD3} studied the convergence of BTM for European and American path dependent options by PDE approach. Liang  et al \cite{LH1} obtained a convergence rate of the BTM for American put options with penalty method and Hu et al \cite{HL} obtained an optimal convergence rate for an explicit finite difference scheme and BTM for a variational inequality problem of American put options.

BTM is extended to the jump-diffusion models for option pricing.  Amin \cite{ami} generalized their algorithm of \cite{AK} to jump-diffusion models. Zhang \cite{zha} studied numerical analysis for American option in jump-diffusion models. Xu et al \cite{XQJ} studied numerical analysis for BTM for European options in Amin's jump-diffusion models and gave strict error estimation for explicit difference scheme and optimal error estimation for BTM. Qian et al \cite{QXJ} proved equivalence of BTM and explicit difference scheme for American option in jump-diffusion models, convergence of explicit difference scheme, existence and monotones of optimal exercise boundary. Luo \cite{luo} studied approximated optimal exercise boundary of American option in jump-diffusion model. Liang \cite{lia} obtained a convergence rate of the BTM for American put options in jump-diffusion models. Liang et al \cite{LH2} obtained an optimal convergence rate for BTM for a variational inequality problem of American put options in jump-diffusion models and a convergence rate estimate of approximated optimal exercise boundary to the actual free boundary.

The above all results are obtained under the assumption that the interest rate and volatility are all constants. 

On the other hand, Jiang \cite{jia} studied Black-Scholes PDE with time dependent coefficients as a model for European options in diffusion model and provided the generalized Black-Scholes formula. H.C. O et al \cite{O3} derived a pricing formula of higher order binary with time dependent coefficients and using it, studied the pricing problem of corporate zero coupon bonds. Such higher order binaries with time dependent coefficients are arising in the pricing problem of corporate bonds with discrete coupon (\cite{O2}). H.C. O et al \cite{O1} studied some general properties of solutions to inhomogeneous Black-Scholes PDEs with discontinuous maturity payoffs and time dependent coefficients.

This article concerns with binomial tree methods and monotonic properties for American put options with time dependent coefficients. We consider monotonic properties and convergences of prices by binomial tree methods and explicit difference schemes for the variational inequality model of American put options with time dependent coefficients and then using them prove the decreasing property of the price of American put options and increasing property of the optimal exercise boundary on time variable.

When the coefficients are time dependent, in particular, in the case with time dependent volatility, it is not reasonable to assume that the dynamics of the underlying asset’s price forms a binomial tree if we use a partition of time interval with equal parts. Thus one of our main problems is to find a time interval partition method that allows binomial tree dynamics of the underlying asset’s price. Another point is to prove the monotonic property of option price and approximated optimal exercise boundary. Jiang and Dai’s convergence proof (\cite{JD2}) strongly depends on the monotonic property of option price but such monotonic property of option price may not hold when coefficients including interest rate and volatility are time dependent as you can see in the following remark 3.2. We found a special time interval partition method and conditions under which the prices of American put option by BTM and explicit difference scheme have the monotonic property on time variable. Such a special partition of time interval needs some annoying consideration in proving convergence to viscosity solutions.  

The remainder of this article is organized as follows. In section 2, we find a time interval partition method that allows binomial tree dynamics of the underlying asset’s price and briefly mention BTM for European options. In section 3, we study BTM price of American put option, its monotonic property and existence of approximate optimal exercise boundary. In section 4, we study explicit difference scheme for variational inequality model for American put option and show the monotonicity of option price on time-variable and existence of approximated optimal exercise boundary. Section 5 is devoted to the convergence proof of the explicit difference scheme and BTM.  In Section 6, Section 7 and Section 8 the linear homogeneity and call-put symmetry of the price functions in the BTM and the EDS for the variational inequality model of American options with time dependent coefficients are studied and the results on American call options are provided.  

%
%

\section{Time Interval Partition and BTM for European Options with Time dependent coefficients.}

Let $r(t)$, $q(t)$ and $\sigma(t)$ be the interest rate, the dividend rate and the volatility of the underlying asset of option, respectively. Let $0=t_0<t_1<\cdots<t_N=T$ be a partition of life time interval $[0,T]$ and denote as follows
$$r_n=r(t_n), q_n=q(t_n), ~~~\sigma_n=\sigma(t_n), $$
$$\eta_n=1+q_n\Delta t_n,  ~~~\rho_n=1+r_n\Delta t_n,$$
$$\Delta t_n=t_{n+1}-t_n, ~~~~n=1,\cdots,N-1.$$
The volatility $\sigma(t)$  of the underlying asset determines the fluctuation of its price in time interval $[t,t+\Delta t]$. So if we divide $[0,T]$  by equal parts, then the dynamics of the underlying asset’s price in subinterval $[t_n,t_{n+1}]$  of time may not form a binomial tree. It makes BTM difficult in the case with time dependent coefficients . 

On the other hand, from the practical meaning of the volatility $\sigma_n $, although we consider the underlying asset’s price $S$ in the some interval $[t_n,t_{n+1}]$ , the underlying asset’s price $S$ largely changes if $\sigma_n $ is large; the underlying asset’s price $S $ changes a little if $\sigma_n $ is small.  So we can imagine that we can make the widths of changes of $S$ in all subintervals a constant if we differently define the length $\Delta t_n=t_{n+1}-t_n$  of subinterval  $[t_n,t_{n+1}]$ according to the size of $\sigma_n $. In other words, if we define $\Delta t_n=t_{n+1}-t_n$  such that $\sigma_n\cdot \Delta t_n=const=(\ln u)^2$ , then we can assume that the width of change of $S$ in every subinterval $[t_n,t_{n+1}]$  is $u$ and the dynamics of $S$ in every subinterval $[t_n,t_{n+1}]$  satisfies one period - two states model \cite{jia}. Then $S_t$ are random variables and the evolution in $[0,T]$ forms a binomial tree. Such a partition method provides a key to overcome the difficulty arising in the case with time dependent coefficients.

Let us define $t_n(n=1,\cdots,N)$ more definitely. Let assume $u>1$. First, we define
$$t_0=0, ~~\sigma_0=\sigma(t_0), ~~\Delta t_0=\frac{(\ln u)^2}{\sigma_0^2},~~ t_1=t_0+\Delta t_0=(\ln u)^2\cdot\frac{1}{\sigma_0^2}.$$
If $t_1\le T$, then we define as follows:
$$\sigma_1=\sigma(t_1), ~~\Delta t_1=\frac{(\ln u)^2}{\sigma_1^2},~~ t_2=t_1+\Delta t_1=(\ln u)^2\cdot\left(\frac{1}{\sigma_0^2}+\frac{1}{\sigma_1^2}\right).$$
Inductively, if $t_n\le T$, then we define as follows:
\begin{equation} \label{eq2-1}   
\sigma_n=\sigma(t_n), ~~\Delta t_n=\frac{(\ln u)^2}{\sigma_n^2},~~ t_{n+1}=t_n+\Delta t_n=(\ln u)^2\cdot\left(\frac{1}{\sigma_0^2}+\cdots+\frac{1}{\sigma_n^2}\right).
\end{equation} 
Such a process is continued until $t_N\le T<t_{N+1}$. Then the number $N$ of subintervals depends on $u,T$  and $\sigma(t)$.

If we assume that 
\begin{equation} \label{eq2-2}   
0<\underline{\sigma}\le \sigma(t)\le \bar{\sigma},
\end{equation} 
then we obtain lower and upper bounds for the size $\Delta t_n$ of subintervals of time and the number $N$ of subintervals. From the definition \eqref{eq2-1} of  $\Delta t_n$ , we have 
\begin{equation} \label{eq2-3}   
\frac{(\ln u)^2}{\bar{\sigma}^2}\le \Delta t_n\le \frac{(\ln u)^2}{\underline{\sigma}^2}.
\end{equation} 

On the other hand, if we use $t_N\le T<t_{N+1}$, then we have
\begin{equation} \label{eq2-4}   
\frac{T\cdot\underline{\sigma}^2}{(\ln u)^2}-1< N \le \frac{T\cdot\bar{\sigma}^2}{(\ln u)^2}.
\end{equation} 

{\bf Remark 2.1}. 
If $u\downarrow 1$, then $N\rightarrow \infty$  and $0\le T-t_N<\Delta t_N=(\ln u)^2\cdot \frac{1}{\sigma_N^2}\rightarrow 0$ .\\

Now we consider the dynamics of the underlying asset's price S. Assume that the width of change of $S$ in every subinterval $[t_n,t_{n+1}]$ is $u$ , $d=u^{-1}$  and the dynamics of $S$ in every subinterval $[t_n,t_{n+1}]$ satisfies one period - two states model. That is, the underlying asset’s price $S_{t_n}$ at time $t_n$ is changed into $S_{t_n}u$ or $S_{t_n}d$ . If the initial price of $S$ is $S_0$ , then $S_{t_n}$ can take one of the following values
$$ S_\alpha^n=S_0u^{n-\alpha}d^\alpha~(0\le \alpha\le n)  ~~\text{or}~~  S_j=S_0u^j~(j=n,n-2,\cdots, -n+2, -n). $$

Assume that
\begin{equation} \label{eq2-5}   
d \eta_n < \rho_n < u\eta_n,~n=0,1,\cdots,N.
\end{equation} 
If we denote
\begin{equation} \label{eq2-6}     
\theta_n=\frac{\rho_n/\eta_n-d}{u-d},~n=0,1,\cdots,N.
\end{equation} 
then we have $0<\theta_n<1$  and BTM price of European option with time dependent coefficients is provided as follows:
\begin{eqnarray}\label{eq2-7}   
& & V_\alpha^N=(S_\alpha^N-E)^+~\text{(for call)  or}~(E-S_\alpha^N)^+~\text{(for put)},~0\le\alpha\le N,\nonumber\\
& & V_\alpha^n=\frac{1}{\eta_n}[\theta_nV_{\alpha+1}^{n+1}+(1-\theta_n)V_{\alpha-1}^{n+1}], 0\le\alpha\le n,~n=N-1,\cdots,~1,~0.   
\end{eqnarray} 

{\bf Remark 2.2}. 
Using Jiang's method (\cite{jia}), we can easily prove the followings: BTM can be seen as a special explicit difference scheme for Black-Scholes PDE 
\begin{eqnarray}\label{eq2-8}    
& & \frac{\partial V}{\partial t}+\frac{\sigma^2(t)}{2}\frac{\partial^2 V}{\partial x^2} 
+\left[r(t)-q(t)-\frac{\sigma^2(t)}{2} \right]\frac{\partial V}{\partial x}-r(t)V=0, ~-\infty<x<\infty,~0\le t<T,\nonumber\\
& & V(x,T)=(e^x-E)^+~\text{or}~(E-e^x)^+, ~-\infty<x<\infty.
\end{eqnarray} 
Let $x_m=m\Delta x~(-\infty<m<\infty)$, $0=t_0<t_1<\cdots<t_N=T$ and $\Delta t_n=t_{n+1}-t_n$. Denote $V_m^n=V(x_m,~t_n)$ . Then the explicit difference scheme for \eqref{eq2-8} is provided as follows: 
\begin{eqnarray}\label{eq2-9}   
& & V_m^N=(e^{m\Delta x}-E)^+~\text{or}~(E-e^{m\Delta x})^+~,\nonumber\\
& & V_m^n=\frac{1}{1+r_n\Delta t_n}\left\{\left(1-\frac{\sigma_n^2\Delta t_n}{\Delta x^2}\right)V_m^{n+1}+\left[\frac{\sigma_n^2\Delta t_n}{2\Delta x^2}+\frac{1}{2}\left(r_n-q_n-\frac{\sigma_n^2}{2} \right)\frac{\Delta t_n}{\Delta x}\right]V_{m+1}^{n+1}\right. \nonumber\\
& & ~~~~~\left.+\left[\frac{\sigma_n^2\Delta t_n}{2\Delta x^2}-\frac{1}{2}\left(r_n-q_n-\frac{\sigma_n^2}{2} \right)\frac{\Delta t_n}{\Delta x}\right]V_{m-1}^{n+1}\right\}, ~n=N-1,\cdots,~1,~0.
\end{eqnarray} 
The scheme \eqref{eq2-9} is consistent if $r(t)$ ,$q(t)$ and $\sigma(t)$ are bounded and continuous on $[0,T]$. Such an explicit difference scheme is stable if 
$$\sigma_n^2\Delta t_n\le\Delta x^2~;~~1-\frac{1}{\sigma_n^2}\left|r_n-q_n-\frac{\sigma_n^2}{2}\right|\Delta x\ge 0,~~0\le \forall n \le N-1.$$
Let $\Delta x \rightarrow 0$ . Then $\Delta t_n \rightarrow 0$ and \eqref{eq2-9} converges to the solution to \eqref{eq2-8}. So BTM price \eqref{eq2-7} also converges to the solution to \eqref{eq2-8}.


\section{BTM for American Put Options with Time Dependent Coefficients.}

Let $0=t_0<t_1<\cdots<t_N\le T$ be the partition of time defined in \eqref{eq2-1} and let
$$S_j=S_0u^j~(j=n,n-2,\cdots,-n+2, -n~;~n=0,\cdots,N), $$
$$ \varphi_j=(E-S_j)^+.$$
Then BTM prices $V_j^n=V(S_j,t_n)$  of American put option are provided as follows: 
\begin{eqnarray}\label{eq3-1}     
& & V_j^N=\varphi_j,\nonumber\\
& & V_j^n=\max\left\{\frac{1}{\rho_n}\left[\theta_nV_{j+1}^{n+1}+(1-\theta_n)V_{j-1}^{n+1}\right], \varphi_j\right\}, ~n=N-1,\cdots,~1,~0.
\end{eqnarray} 

Now we consider the monotonic property of BTM price $V_j^n$ for American put option.

\begin{theorem} \label{theorem3-1}
 BTM prices of American put option
\begin{equation} \label{eq3-2}       
 V_j^n=P(S_j, t_n~;~E)~(n=0,~1,\cdots,N~,j=n,~n-2,\cdots, -n+2, -n)
\end{equation} 
are decreasing with respect to $S_j$ and increasing with respect to  $E$. That is, 
\begin{eqnarray*}
 & & V_j^n=P(S_j, t_n~;~E)\ge P(S_{j+1}, t_n~;~E)=V_{j+1}^n,\\
 & & P(S_j, t_n~;~E_1)\le P(S_{j}, t_n~;~E_2)~~\text{if}~~E_1<E_2.
\end{eqnarray*} 
\end{theorem} 
\begin{proof}  $V_j^N=\varphi_j=(E-S_j)^+$ is decreasing function on $S_j$ and increasing on $E$. Now assume that   $ V_j^{k+1}\ge V_{j+1}^{k+1}$ when $n=k+1$ . Then we have
\begin{eqnarray*}   
V_j^k&=&\max\left\{\frac{1}{\rho_k}\left[\theta_kV_{j+1}^{k+1}+(1-\theta_k)V_{j-1}^{k+1}\right], \varphi_j\right\}\\
&\ge&\max\left\{\frac{1}{\rho_k}\left[\theta_kV_{j+2}^{k+1}+(1-\theta_k)V_{j}^{k+1}\right], \varphi_{j+1}\right\}=V_{j+1}^k.
\end{eqnarray*} 
Thus $V_j^k$ is decreasing on $S_j$. Similarly, we can prove $V_j^k$ is increasing on $E$. (QED) 
\end{proof}
\\

In order to prove that $ V_j^n$ is decreasing on time variable, we need the following lemmas.
\begin{lemma} \label{lemma3-1}
(i) {\it If $r(t)/\sigma^2(t)$ is increasing on $t$, then $\rho_n\le \rho_{n+1}.$}\\
(ii) {\it If $q(t)/\sigma^2(t)$ is decreasing on $t$, then $\eta_n\ge \eta_{n+1}.$}\\
(iii) {\it If $r(t)/\sigma^2(t)$ is increasing, $q(t)/\sigma^2(t)$ decreasing and $\Delta t_n$ is sufficiently small, then} 
$$\rho_n/\eta_n\le \rho_{n+1}/\eta_{n+1}~;~\theta_n\le\theta_{n+1}.$$
\end{lemma} 
\begin{proof} (i) If $r(t)/\sigma^2(t)$ is increasing, then from the definition of $\Delta t_n$,  we have
\begin{eqnarray*}   
& & \frac{r_{n+1}}{\sigma^2_{n+1}}\ge\frac{r_{n}}{\sigma^2_{n}}\Leftrightarrow (\ln u)^2\cdot\frac{r_{n+1}}{\sigma^2_{n+1}}\ge(\ln u)^2\cdot\frac{r_{n}}{\sigma^2_{n}}\Leftrightarrow\\
& &\rho_{n+1}=1+r_{n+1}\Delta t_{n+1}\ge1+r_{n}\Delta t_n=\rho_n.
\end{eqnarray*} 
(ii) is proved in similar way with (i).\\
(iii) If $\Delta t_n$ is sufficiently small, then $\eta_n>0$ . Since  $\rho_n\le \rho_{n+1}$ and $\eta_n\ge \eta_{n+1}$, we have $\rho_n/\eta_n\le \rho_{n+1}/\eta_{n+1}$. Thus from \eqref{eq2-6}, we have $\theta_n\le\theta_{n+1}$. (QED)
\end{proof}

\begin{lemma} \label{lemma3-2}
(i) {\it If $A\le B$ and $0\le\alpha\le\beta$, then $\alpha A+(1-\alpha)B\ge\beta A+(1-\beta)B.$}
\end{lemma} 
\begin{proof} $\alpha A+(1-\alpha)B-\beta A-(1-\beta)B=(\beta-\alpha)(B-A)\ge 0$.
 (QED)
\end{proof}
\begin{theorem}  \label{theorem3-2}
Assume that \eqref{eq2-5} is satisfies, $r(t)/\sigma^2(t)$ is increasing and $q(t)/\sigma^2(t)$ decreasing on $t$. Then for BTM prices $V_j^n$ of American put option we have
$$ V_j^{n-1}\ge V_j^{n}.$$ 
\end{theorem} 
\begin{proof}  From \eqref{eq3-1} we have
 $$V_j^{N-1}\ge\varphi_j=V_j^N~(j=N,N-2,\cdots,-N+2,-N).$$ 
Now assume that   $ V_j^{k}\ge V_j^{k+1}~(\forall j)$. Then we have
\begin{eqnarray*}   
V_j^{k-1}&=&\max\left\{\frac{1}{\rho_{k-1}}\left[\theta_{k-1}V_{j+1}^{k}+(1-\theta_{k-1})V_{j-1}^{k}\right], \varphi_j\right\}\\
&\ge&\max\left\{\frac{1}{\rho_{k-1}}\left[\theta_{k-1}V_{j+1}^{k+1}+(1-\theta_{k-1})V_{j-1}^{k+1}\right], \varphi_j\right\}\\
&\ge&\max\left\{\frac{1}{\rho_{k}}\left[\theta_{k-1}V_{j+1}^{k+1}+(1-\theta_{k-1})V_{j-1}^{k+1}\right], \varphi_j\right\}\\
&\ge&\max\left\{\frac{1}{\rho_k}\left[\theta_k V_{j+1}^{k+1}+(1-\theta_k)V_{j-1}^{k+1}\right], \varphi_{j+1}\right\}=V_{j}^k.
\end{eqnarray*} 
Here the first inequality comes from the induction assumption $ V_j^{k}\ge V_j^{k+1}~(\forall j)$ , the second inequality from lemma \ref{lemma3-1} (i), the last inequality from lemma \ref{lemma3-1} (iii), theorem \ref{theorem3-1} and lemma \ref{lemma3-2}. (QED)
\end{proof}\\

{\bf Remark 3.1}. 
Theorem \ref{theorem3-2} strongly represents the effect of time dependent coefficients. Here the main tools are lemma \ref{lemma3-1} and lemma \ref{lemma3-2}.\\ 

{\bf Remark 3.2}.
The conditions of theorem \ref{theorem3-2} are essential. See the following figures:
\begin{figure}[tbhp]
\centering
\includegraphics[width=12.5cm , height=9cm]{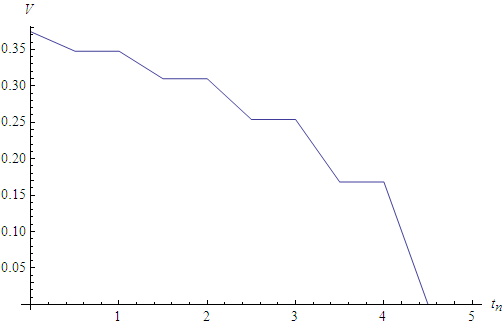}
\caption{Plot $(t_n : V(S_j,t_n))$ when $r(t)=0.1,q=0,\sigma=1,T=5, E=1, j=1$}
\end{figure}
\begin{figure}[tbhp]
\centering
\includegraphics[width=12.5cm , height=7.5cm]{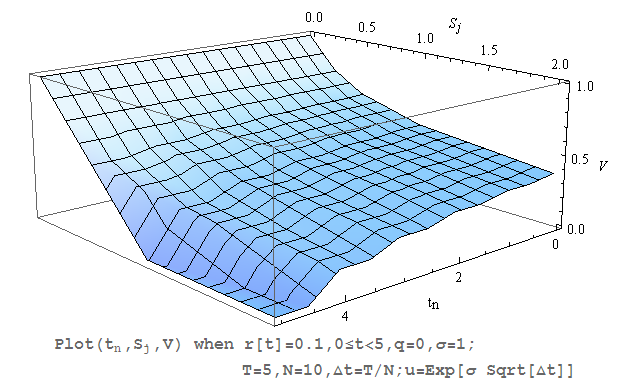}
\caption{$r(t)$ is increasing, so $V$ is decreasing on $t$.}
\end{figure}
\begin{figure}[tbhp]
\centering
\includegraphics[width=10.5cm , height=6.5cm]{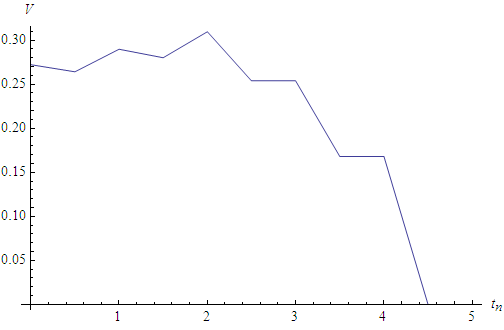}
\caption{Plot $(t_n : V(S_j,t_n))$ when $r(t)=Piecewise\{\{0.2,0\le t<2\},\{0.1,2\le t<5\}\}; q=0,\sigma=1,T=5, E=1, j=1$}
\end{figure}
\begin{figure}[tbhp]
\centering
\includegraphics[width=14cm , height=7cm]{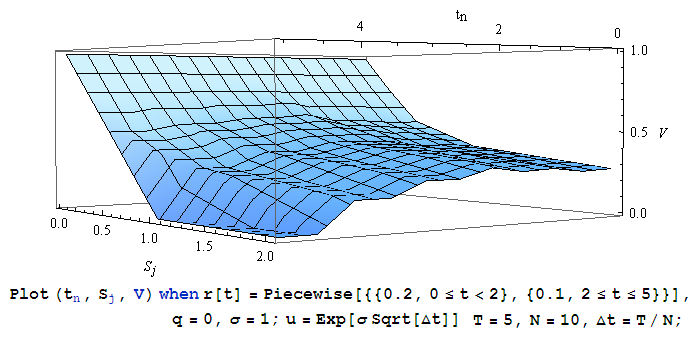}
\caption{$r(t)$ is not increasing, so $V$ is not decreasing on $t$}
\end{figure}

{\bf Remark 3.3}. Only using the analogs of lemma \ref{lemma3-1} and lemma \ref{lemma3-2}, it seems difficult to prove that American call option's BTM price is decreasing on $t$. \\

Now we consider the existence of approximated optimal exercise boundary.

\begin{theorem}  \label{theorem3-3}
Let $\Delta t_n$ be sufficiently small. Under the conditions of theorem \ref{theorem3-2} , for every $t_n~(0\le n\le N-1)$, there exists a $j_n\in Z$ such that
\begin{eqnarray} \label{eq3-3}     
&&V_j^n=\varphi_j~\text{for}~j\le j_n, \nonumber\\
&&V_j^n>\varphi_j~\text{for}~j=j_n+1, \nonumber\\
&&V_j^n\le\varphi_j~\text{for}~j\ge j_n+2.
\end{eqnarray} 
Furthermore we have 
\begin{eqnarray} \label{eq3-4}  
j_{n-1}\le j_n.
\end{eqnarray} 
\end{theorem} 
\begin{proof}   
Without loss of generality, we assume that $S_0=1$ and $E=1$. (Otherwise, use change of variables $\hat{S}=S/E,~\hat{h}=V/E$.) Since
 $$V_j^{N}=(1-S_j)^+=\varphi_j,~(j=N,N-2,\cdots,-N+2,-N),$$ 
we have 
$$\varphi_j=V_j^{N}=0~(j\ge 0);~\varphi_j=V_j^{N}>0,~(j\le -1).$$ 
Since $V_j^{N-1}=\max\left\{\frac{1}{\rho_{N-1}}\left[\theta_{N-1}\varphi_{j+1}+(1-\theta_{N-1})\varphi_{j-1}\right], \varphi_j\right\}$, we have
$$V_j^{N-1}\ge 0=V_j^N=\varphi_j,~j\ge 0. $$
In particular, 
\begin{equation} \label{eq3-5}   
V_j^{N-1}=\varphi_j~(j\ge 1);~ V_0^{N-1}=\rho_{N-1}^{-1}(1-\theta_{N-1})\varphi_{-1}>0=\varphi_{0}. 
\end{equation}
Now we consider the case of $j\le -1$. 
\begin{eqnarray} \label{eq3-6}   
V_j^{N-1}&=&\max\left\{\frac{1}{\rho_{N-1}}\left[\theta_{N-1}\varphi_{j+1}+(1-\theta_{N-1})\varphi_{j-1}\right], \varphi_j\right\}\nonumber\\
&=&\max\left\{\frac{1}{\rho_{N-1}}\left[\theta_{N-1}(1-u^{j+1})+(1-\theta_{N-1})(1-u^{j-1})\right], \varphi_j\right\}\nonumber\\
&=&\max\left\{\rho_{N-1}^{-1}-\eta_{N-1}^{-1}u^j,~1-u^j\right\}.
\end{eqnarray} 
Note that if $j\rightarrow -\infty$ , then $\rho_{N-1}^{-1}-\eta_{N-1}^{-1}u^j<1$ and $1-u^j\rightarrow 1$ . So there exists  
$$j_{N-1}=\max\{j\in Z:j\le -1, \rho_{N-1}^{-1}-\eta_{N-1}^{-1}u^j\le 1-u^j\}. $$
If $j\le j_{N-1}$ , then we have $\rho_{N-1}^{-1}-\eta_{N-1}^{-1}u^j\le 1-u^j$ and thus $V_j^{N-1}=1-u^j=\varphi_j$. If $j_{N-1}+1\le j\le -1$ , then we have  $\rho_{N-1}^{-1}-\eta_{N-1}^{-1}u^j>1-u^j$ and thus $V_j^{N-1}>\varphi_j$ . So $j_{N-1}$  satisfies \eqref{eq3-3} with $n=N-1$ . (In particular if $\eta_{N-1}\le 1(\Leftrightarrow q_{N-1}\le 0)$ , then $j_{N-1}=-1$.)

Now assume that when $n=k$ , there exists $j_k$ satisfying \eqref{eq3-3} and \eqref{eq3-4}. Then if $j\le j_k-1$, then $V_{j-1}^k=\varphi_{j-1},~V_{j+1}^k=\varphi_{j+1}$ and thus from the formula \eqref{eq3-1} and the same calculation in \eqref{eq3-6} we have 
$$V_j^{k-1}=\max\left\{\frac{1}{\rho_{k-1}}\left[\theta_{k-1}\varphi_{j+1}+(1-\theta_{k-1})\varphi_{j-1}\right], \varphi_j\right\}=\max\left\{\frac{1}{\rho_{k-1}}-\frac{u^j}{\eta_{k-1}},~1-u^j\right\}.$$
In the case that $\eta_{k-1}>1~(\Leftrightarrow q_{k-1}> 0)$  , let
$$l=\max\{j\in Z:j\le j_k-1, \rho_{k-1}^{-1}-\eta_{k-1}^{-1}u^j\le 1-u^j\}, $$
then we have $l\le j_k-1$ . If $l<j_k-1$ , then we define $j_{k-1}=l$ . Then using the similar way with the consideration when $n=N-1$ and theorem \ref{theorem3-2}, we have
$$j\le j_{k-1}\Rightarrow V_j^{k-1}=\varphi_j;~j=j_{k-1}+1\Rightarrow V_j^{k-1}>\varphi_j;~j\ge j_{k-1}+2\Rightarrow V_j^{k-1}\ge V_j^{k}\ge\varphi_j.$$
If  $l=j_k-1$ (that is, $V_j^{k-1}=\varphi_j$  for all $j\le j_k-1$ ), then note that 
$$j=j_{k-1}+1\Rightarrow V_j^{k-1}\ge V_j^k>\varphi_j;~j\ge j_{k-1}+2\Rightarrow V_j^{k-1}\ge V_j^k\ge\varphi_j.$$
Generally, we have $V_j^k\ge\varphi_j$   when $j=j_k$ . So if $V_{j_k}^k>\varphi_{j_k}$ , then we define $j_{k-1}=j_k-1$. If $V_{j_k}^k=\varphi_{j_k}$ , then we define $j_{k-1}=j_k$ . Thus in any case  $j_{k-1}(\le j_k)$ is well defined. 
(QED)
\end{proof}\\
%
%

\section{The Explicit Difference Scheme for Variational Inequality Model of American Options with Time Dependent Coefficients.}

A Variational Inequality pricing model of American option with time dependent coefficients is provided as follows:
\begin{eqnarray}\label{eq4-1}     
&&\min\left\{-\frac{\partial V}{\partial t}-\frac{\sigma(t)^{2}}{2}S^2\frac{\partial^2 V}{\partial S^2}-(r(t)-q(t))S\frac{\partial V}{\partial S}+r(t)V,V-\psi\right\}=0,\nonumber\\&&\quad\quad\quad\quad\quad\quad\quad\quad\quad 0\leq t<T,~0<S<\infty,\nonumber\\
&&V(S,T)=\psi(S),\quad~~0<S<\infty.     
\end{eqnarray}
Here 
$$\psi(S)=(S-E)^+~~\text{(for call)},\quad \psi(S)=(E-S)^+~~\text{(for put).}$$
Using the transformation 
\begin{eqnarray}\label{eq4-2}     
u(x,t)=V(S,t);~~S=e^x,     
\end{eqnarray}
the problem \eqref{eq4-1} is changed to the following problem
\begin{eqnarray}\label{eq4-3}     
&&\min\left\{-\frac{\partial u}{\partial t}-\frac{\sigma(t)^{2}}{2}\frac{\partial^2 u}{\partial x^2}-\left(r(t)-q(t)-\frac{\sigma(t)^{2}}{2}\right)\frac{\partial u}{\partial x}+r(t)u,u-\varphi\right\}=0,\nonumber\\&&\quad\quad\quad\quad\quad\quad\quad\quad\quad 0\leq t<T,~-\infty<x<\infty,\nonumber\\
&&u(x,T)=\varphi(x),\quad~~-\infty<x<\infty.     
\end{eqnarray}
Here  
$$\varphi(x)=(e^x-E)^+~~\text{(for call)},\quad \varphi(x)=(E-e^x)^+~~\text{(for put).}$$

We construct a lattice on $\Sigma = \{-\infty<x<\infty, 0\le t<T\}$ as follows: Select any $c\in R$ and $\Delta x$ . Let $x_j=j\Delta x+c$ . When $0<\alpha\le 1$ , we define as follows: 
$$t_0=0, ~~\Delta t_0=\frac{\alpha\Delta x^2}{\sigma^2(t_0)},~~ t_1=t_0+\Delta t_0, ~~\Delta t_1=\frac{\alpha\Delta x^2}{\sigma^2(t_1)},\cdots,$$
\begin{equation} \label{eq4-4}        
t_{n}=t_{n-1}+\Delta t_{n-1}, ~~\Delta t_n=\frac{\alpha\Delta x^2}{\sigma^2(t_n)},~~ n=0,~1,~2,\cdots.
\end{equation} 
This process is continued until  $t_N$ such that
$$t_N=t_{N-1}+\Delta t_{N-1}\le T<t_{N+1}=t_{N}+\frac{\alpha\Delta x^2}{\sigma^2(t_N)}.$$
Then we have a lattice on $\Sigma = \{-\infty<x<\infty, 0\le t<T\}$:
\begin{equation} \label{eq4-5}        
Q_c=\{(x_j,t_{n}):x_j=j\Delta x+c, 0\le n\le N, j\in Z\}.
\end{equation} 
Under the assumption \eqref{eq2-2} we have
\begin{equation} \label{eq4-6}   
\frac{\alpha\Delta x^2}{\bar{\sigma}^2}\le \Delta t_n\le \frac{\alpha\Delta x^2}{\underline{\sigma}^2}.
\end{equation} 
Thus there exists $N$ such that  $t_N\le T<t_{N+1}$ and we have
\begin{equation} \label{eq4-7}   
\frac{T\cdot\underline{\sigma}^2}{\alpha\Delta x^2}-1< N \le \frac{T\cdot\bar{\sigma}^2}{\alpha\Delta x^2}.
\end{equation} 
Therefore if $\Delta x\rightarrow 0$, then $N\rightarrow \infty$  and $0\le T-t_N\le \Delta t_N\le  \frac{\alpha\Delta x^2}{\underline{\sigma}^2}\rightarrow 0$ .

$u_j^n=u(j\Delta x+c,~t_n)$ represents the value of approximation at  $(j\Delta x+c,~t_n)$ and let $\varphi_j=\varphi(j\Delta x+c)$. Taking explicit difference for time and the conventional difference discretization for space variable in \eqref{eq4-3}, we have
\begin{eqnarray}\label{eq4-8}     
&&\min\left\{-\frac{u_j^{n+1}-u_j^{n}}{\Delta t_n}-\frac{\sigma^{2}(t_n)}{2}\cdot\frac{u_{j+1}^{n+1}-2 u_j^{n+1}+u_{j-1}^{n+1}}{\Delta x^2}\right.\nonumber\\
&&\quad\quad\left.-\left[r(t_n)-q(t_n)-\frac{\sigma^{2}(t_n)}{2}\right]\frac{u_{j+1}^{n+1}-u_{j-1}^{n+1}}{2\Delta x}+r(t_n)u_j^n,~u_j^n-\varphi_j\right\}=0.
\end{eqnarray}
If we denote $r_n=r(t_n), q_n=q(t_n), \sigma_n=\sigma(t_n)$ , then \eqref{eq4-8} is equivalent to
\begin{eqnarray*}
~u_j^n=\max\left\{\frac{1}{1+r_n\Delta t_n}\left\{\left(1-\frac{\sigma^{2}_n\Delta t_n}{\Delta x^2}\right)u_j^{n+1}+\frac{\sigma^{2}_n\Delta t_n}{\Delta x^2}\left[\left(\frac{1}{2}+\frac{\Delta x}{2\sigma^{2}_n}\left(r_n-q_n-\frac{\sigma^{2}_n}{2}\right)\right)u_{j+1}^{n+1}\right.\right.\right.\nonumber\\
\left.\left.\left.+\left(\frac{1}{2}-\frac{\Delta x}{2\sigma^{2}_n}\left(r_n-q_n-\frac{\sigma^{2}_n}{2}\right)\right)u_{j-1}^{n+1}\right]\right\},~\varphi_j\right\}.\quad\quad\quad\quad\quad\quad\quad\quad\quad\quad\quad\quad\quad\quad
\end{eqnarray*}
Here,  if we denote $S_0=e^c$ , then 
$$\varphi_j=(S_0e^{j\Delta x}-E)^+~~\text{(for call)},\quad \varphi_j=(E-S_0e^{j\Delta x})^+~~\text{(for put).}$$
From \eqref{eq4-4} we have $\alpha=\frac{\sigma^{2}_n\Delta t_n}{\Delta x^2}$  and let 
\begin{eqnarray}\label{eq4-9} 
a_n=\frac{1}{2}+\frac{\Delta x}{2\sigma^{2}_n}\left(r_n-q_n-\frac{\sigma^{2}_n}{2}\right).
\end{eqnarray}
Then we have the explicit difference scheme
\begin{eqnarray}
&&U_j^N=\varphi_j,~j\in Z,        \label{eq4-10}  \\
&&U_j^n=\max\left\{\frac{1}{\rho_n}\left\{(1-\alpha)U_j^{n+1}+\alpha\left[a_n U_{j+1}^{n+1}+(1-a_n)U_{j-1}^{n+1}\right]\right\},~\varphi_j\right\}, \label{eq4-11}\\
&&\quad\quad\quad\quad\quad\quad\quad\quad\quad\quad\quad\quad\quad\quad\quad\quad\quad\quad\quad\quad\quad\quad n=N-1,\cdots,~1,~0.\nonumber
\end{eqnarray}
(Note that $\rho_n=1+r_n\Delta t_n$.) In particular, if $\alpha=\frac{\sigma^{2}_n\Delta t_n}{\Delta x^2}=1$, then $\Delta x=\sigma_n\sqrt{\Delta t_n}$ and
\begin{eqnarray}\label{eq4-12} 
U_j^n=\max\left\{\frac{1}{\rho_n}\left[a_n U_{j+1}^{n+1}+(1-a_n)U_{j-1}^{n+1}\right],~\varphi_j\right\}, \quad n=N-1,\cdots,~1,~0.
\end{eqnarray}\\

Now we consider the relation of BTM and explicit difference scheme for American option.\begin{lemma} \label{lemma4-1}
{\it If ~ $\ln u=\Delta x=\sigma_n\sqrt{\Delta t_n}$,~we have }  
$$\theta_n=\frac{1}{2}+\frac{\Delta x}{2\sigma^{2}_n}\left(r_n-q_n-\frac{\sigma^{2}_n}{2}\right)+O(\Delta x^3).$$
{\it Here $\theta_n$ are coefficients of BTM defined by} \eqref{eq2-6}. 
\end{lemma} 

The proof  is easy.\\

Contrasting \eqref{eq3-1} and \eqref{eq4-12}, BTM is equivalent to a special explicit difference scheme \eqref{eq4-12} in the sense of neglecting $O(\Delta x^3)$.\\

Now we show the conditions for American (put) option price be monotonic. 

\begin{theorem} \label{theorem4-1}
 Assume that $0<\alpha\le 1$ ~and~ $\left|\frac{\Delta x}{\sigma^{2}_n}\left(r_n-q_n-\frac{\sigma^{2}_n}{2}\right)\right|<1$.\\  
(i) If $\varphi_j=(S_0e^{j\Delta x}-E)^+$ (call), then  $U_j^n\le U_{j+1}^n$ and $0\le U_j^n\le e^{j\Delta x+c}$.\\
(ii) If $\varphi_j=(E-S_0e^{j\Delta x})^+$ (put), then   $ U_j^n\ge U_{j+1}^n$ and $0\le U_j^n\le E$.
\end{theorem} 
\begin{proof}
From the assumption we have $0<a_n\le 1$.\\
(i) $U_j^N=\varphi_j=(S_0e^{j\Delta x}-E)^+\le (S_0e^{(j+1)\Delta x}-E)^+=\varphi_{j+1}=U_{j+1}^N$.
Now assume that $U_j^{k+1}\le U_{j+1}^{k+1}$. Then we have
\begin{eqnarray*}
U_j^k&=&\max\left\{\frac{1}{\rho_k}\left\{(1-\alpha)U_j^{k+1}+\alpha\left[a_k U_{j+1}^{k+1}+(1-a_k)U_{j-1}^{k+1}\right]\right\},~\varphi_j\right\}\\
&\le&\max\left\{\frac{1}{\rho_k}\left\{(1-\alpha)U_{j+1}^{k+1}+\alpha\left[a_k U_{j+2}^{k+1}+(1-a_k)U_{j}^{k+1}\right]\right\},~\varphi_{j+1}\right\}=U_{j+1}^k.
\end{eqnarray*}
(ii) is proved in the same way as (i). (QED)
\end{proof}

\begin{theorem} \label{theorem4-2}
 Assume that $0<\alpha\le 1$, $\left|\frac{\Delta x}{\sigma^{2}_n}\left(r_n-q_n-\frac{\sigma^{2}_n}{2}\right)\right|<1$, $r(t)/\sigma^2(t)$ is increasing and $q(t)/\sigma^2(t)$ decreasing on $t$. Then prices $U_j^n$ of American put option given by \eqref{eq4-10} and \eqref{eq4-11} with $\varphi_j=(E-S_0e^{j\Delta x})^+$ are decreasing on $t$, that is,  
$$U_j^n\ge U_j^{n+1},~j\in Z,~n=N-1,\cdots,~1,~0.$$
\end{theorem} 
\begin{proof}
When $n=N-1$ , from \eqref{eq4-11} we have $U_j^{N-1}\ge \varphi_j=U_j^N,~j\in Z$ . Assume that $U_j^{k+1}\ge U_j^{k+2}, ~j\in Z$ . From the assumption and lemma \ref{lemma3-1} (i) we have $\rho_k\le\rho_{k+1}$ and thus
\begin{eqnarray*}
U_j^k&=&\max\left\{\frac{1}{\rho_k}\left\{(1-\alpha)U_j^{k+1}+\alpha\left[a_k U_{j+1}^{k+1}+(1-a_k)U_{j-1}^{k+1}\right]\right\},~\varphi_j\right\}\\
&\ge&\max\left\{\frac{1}{\rho_{k+1}}\left\{(1-\alpha)U_{j}^{k+2}+\alpha\left[a_k U_{j+1}^{k+2}+(1-a_k)U_{j-1}^{k+2}\right]\right\},~\varphi_{j}\right\}.
\end{eqnarray*}
From $a_k=\frac{1}{2}+\frac{\Delta x}{2}\left(\frac{r_k}{\sigma^{2}_k}-\frac{q_k}{\sigma^{2}_k}-\frac{1}{2}\right)$ and the assumption, we have $a_k\le a_{k+1}$ . By theorem \ref{theorem4-1} (ii), we have $U_{j-1}^{k+2}\ge U_{j+1}^{k+2}$  and thus lemma \ref{lemma3-2} with $a_k\le a_{k+1}$  gives us
$$a_{k+1} U_{j+1}^{k+2}+(1-a_{k+1})U_{j-1}^{k+2}\le a_k U_{j+1}^{k+2}+(1-a_k)U_{j-1}^{k+2}.$$
Therefore we have
$$U_j^k\ge \max\left\{\frac{1}{\rho_{k+1}}\left\{(1-\alpha)U_{j}^{k+2}+\alpha\left[a_{k+1} U_{j+1}^{k+2}+(1-a_{k+1})U_{j-1}^{k+2}\right]\right\},~\varphi_{j}\right\}=U_j^{k+1}.$$
(QED)
\end{proof}

{\bf Remark 4.1}. Theorem \ref{theorem4-2} strongly represents the effect of time dependent coefficients. Here the main tools are lemma \ref{lemma3-1} (i) and lemma \ref{lemma3-2}. The conditions of theorem \ref{theorem4-2} are essential. If $r(t)/\sigma^2(t)$ is not increasing, then the price of American put option by explicit difference scheme might not be decreasing on $t$ as in remark 4. 

{\bf Remark 4.2}. Only using the analogs of lemma \ref{lemma3-1} and lemma \ref{lemma3-2}, it seems difficult to prove that American call option's price is decreasing on $t$. See Section 6 and 7.  \\

Now we show the existence of approximated optimal exercise boundary.
\begin{theorem} \label{theorem4-3}
Under the assumptions of theorem \ref{theorem4-2}, for any $0\le n\le N-1$ , there exists $j_n\in Z$  such that
\begin{equation}\label{eq4-13} 
j\le j_n\Rightarrow U_j^n=\varphi_j;~j=j_n+1\Rightarrow U_j^n>\varphi_j;~ j\ge j_n+2\Rightarrow U_j^n\ge\varphi_j.
\end{equation}
\begin{equation}\label{eq4-14} 
j_0\le j_1\le\cdots\le j_{N-1}.
\end{equation}
\end{theorem} 
\begin{proof}
Note that $U_j^N=(E-S_0e^{j\Delta x})^+=\varphi_j$ is decreasing on $j\in Z$. Let
\begin{equation}\label{eq4-15} 
k_1=\max\{j\in Z; E-S_0e^{j\Delta x}>0\}.
\end{equation}
Then if  $j\le k_1-1$ then $j-1,~j,~j+1\le k_1$ and
$$\varphi_{j-1}=E-S_0e^{j\Delta x-\Delta x},~\varphi_j=E-S_0e^{j\Delta x},~\varphi_{j+1}=E-S_0e^{j\Delta x+\Delta x}>0.$$ 
Let $u=e^{\Delta x},~d=e^{-\Delta x}$ and $\psi_j=\frac{E}{\rho_{N-1}}-S_0u^j\cdot\frac{1-\alpha+\alpha[a_{N-1}u+(1-a_{N-1})d]}{\rho_{N-1}}$, then 
\begin{eqnarray*}
U_j^{N-1}&=&\max\left\{\frac{1}{\rho_{N-1}}\left[(1-\alpha)\varphi_j+\alpha(a_{N-1}\varphi_{j+1}+(1-a_{N-1})\varphi_{j-1})\right], \varphi_j\right\}\\
&=&\max\{\psi_j,~\varphi_j\}.
\end{eqnarray*} 
Then we have 
$$\lim_{j\rightarrow -\infty}\psi_j=\frac{E}{\rho_{N-1}}<E=\lim_{j\rightarrow -\infty}\varphi_j,~(j\le k_1-1) .$$

First, we consider the case that $\psi_j\le\varphi_j~(\forall j\le k_1-1)$. If $j\le k_1-1$ then $U_j^{N-1}=\varphi_j$,  and if $j=k_1+1$  then $\varphi_j=\varphi_{j+1}=0,~\varphi_{j-1}>0$ and thus we have
$$U_j^{N-1}=\max\left\{\frac{\alpha(1-a_{N-1})}{\rho_{N-1}}\varphi_{j-1},~ 0\right\}>0=\varphi_j.$$ 
If $j\ge k_1+2$ , then from \eqref{eq4-11} we have $U_j^{N-1}\ge\varphi_j$ . So if $U_{j_{k_1}}^{N-1}>\varphi_{j_{k_1}}$ , then we define $j_{N-1}=k_1-1$ ; and if $U_{j_{k_1}}^{N-1}=\varphi_{j_{k_1}}$ , then we define $j_{N-1}=k_1$ .

Next, we consider  the case that $\exists j~(j\le k_1-1):~\psi_j>\varphi_j$ . We define $$j_{N-1}=\max\{j<k_1-1:~\psi_j\le\varphi_j\}.$$ 
Then we have
\begin{eqnarray*}
&&j\le j_{N-1}\quad\quad\Rightarrow ~\psi_j\le\varphi_j\Rightarrow U_j^{N-1}=\varphi_j,\\
&&j=j_{N-1}+1~\Rightarrow ~\psi_j>\varphi_j\Rightarrow U_j^{N-1}=\psi_j>\varphi_j,\\
&&j\ge j_{N-1}+2~\Rightarrow ~\quad\quad\quad\quad\quad U_j^{N-1}\ge\varphi_j.
\end{eqnarray*} 
Thus we proved the existence of $j_{N-1}\le k_1$.

Now we assume that when $n=k+1$  there exists $j_{k+1}$ such that 
\begin{eqnarray}\label{eq4-16}   
&&j_{k+1}\le j_{k+2}\le\cdots\le j_{N-1},\nonumber\\
&&j\le j_{k+1}\quad\quad\Rightarrow  U_j^{k+1}=\varphi_j,\nonumber\\
&&j=j_{k+1}+1~\Rightarrow  U_j^{k+1}>\varphi_j,\nonumber\\
&&j\ge j_{k+1}+2~\Rightarrow  U_j^{k+1}\ge\varphi_j.
\end{eqnarray} 
If $j\le j_{k+1}-1$ , then $j+1,~j,~j-1\le j_{k+1}$ and thus $U_i^{k+1}=\varphi_i~(i=j-1,~j,~j+1)$. As the above, let $\psi_j=\frac{E}{\rho_{k}}-S_0u^j\cdot\frac{1-\alpha+\alpha[a_{k}u+(1-a_{k})d]}{\rho_{k}}$. Then by \eqref{eq4-11} we have
\begin{eqnarray*}
U_j^{k}&=&\max\left\{\frac{1}{\rho_{k}}\left[(1-\alpha)\varphi_j+\alpha(a_{k}\varphi_{j+1}+(1-a_{k})\varphi_{j-1})\right], \varphi_j\right\}\\
&=&\max\{\psi_j,~\varphi_j\}.
\end{eqnarray*} 
Note that $\psi_j<\varphi_j$ for sufficiently large $j\in Z$. In the case that $\psi_j\le \varphi_j~(\forall j\le j_{k+1}-1)$, we have $U_j^k=\varphi_j$ for all $j\le j_{k+1}-1$ . From theorem \ref{theorem4-2} and the inductive assumption \eqref{eq4-16} we have the fact that $j=j_{k+1}+1~\Rightarrow  U_j^k\ge U_j^{k+1}>\varphi_j;~~j\ge j_{k+1}+2~\Rightarrow  U_j^{k}\ge U_j^{k+1}\ge\varphi_j.$
Therefore if $U_{j_{k+1}}^k>\varphi_{j_{k+1}}$ , then let $j_k=j_{k+1}-1$ . If $U_{j_{k+1}}^k=\varphi_{j_{k+1}}$ , then let $j_k=j_{k+1}$ . In the case that  $\exists j~(j\le j_{k+1}-1):\psi_j>\varphi_j$, we define $j_k=\max\{j<j_{k+1}-1:\psi_j\le\varphi_j\}$ . Then  
\begin{eqnarray*}
&&j\le j_{k}\quad\quad\Rightarrow  U_j^{k}=\varphi_j,\\
&&j=j_{k}+1~\Rightarrow  U_j^{k}=\psi_j>\varphi_j,\\
&&j\ge j_{k}+2~\Rightarrow  U_j^{k}\ge\varphi_j.
\end{eqnarray*} 
Thus we proved the existence of  $j_k\le j_{k+1}$. (QED)
\end{proof}

{\bf Remark 4.3}.If $\Delta x$ is enough small, then $j_k\in [j_{k+1}-1,~j_{k+1}]$ . \\

Now we estimate the optimal exercise boundary near the maturity. 

In the first part of the proof of theorem \ref{theorem4-3}, we proved the existence of $j_{N-1}$ , the approximated optimal exercise boundary near the maturity. If  $k_1$ is the one defined in \eqref{eq4-15} and $S_0=e^c$ , then $k_1=\max\{j\in Z; E-e^{j\Delta x+c}>0\}$ and for $j\le k_1-1$ , we have $\varphi_j=E-e^{j\Delta x+c}$  and let 
$$\psi_j=\frac{1}{\rho_{N-1}}\left[(1-\alpha)\varphi_j+\alpha(a_{N-1}\varphi_{j+1}+(1-a_{N-1})\varphi_{j-1})\right].$$

In the case that $\psi_j\le\varphi_j~(\forall j\le k_1-1)$ we know $j_{N-1}=k_1-1$  or $k_1$ . Then we have $E-e^{j_{N-1}\Delta x+c}>0$,~$E-e^{(j_{N-1}+2)\Delta x+c}\le 0$  and thus we have
\begin{equation}\label{eq4-17}   
\ln E - 2\Delta x\le j_{N-1}\Delta x+c\le \ln E.
\end{equation}

In the case that $\exists j~(j\le k_1-1):~\psi_j>\varphi_j$, by theorem \ref{theorem4-3}, we have
$$j_{N-1}=\max\{j\le k_1-1:\psi_j-\varphi_j\le 0 \}.$$
By using the definition of $a_n$  and Taylor expansion, we have 
$$a_ne^{\Delta x}+(1-a_n)e^{-\Delta x}=1+\frac{r_n-q_n}{\sigma_n^2}\Delta x^2+O(\Delta x^4).$$
Then for $j\le k_1-1$  we have
\begin{eqnarray*}
&&\psi_j-\varphi_j=\\
&&=\frac{(1-\alpha)(E-e^{j\Delta x+1})+\alpha\left[a_{N-1}(E-e^{(j+1)\Delta x+1})+(1-a_{N-1})(E-e^{(j-1)\Delta x+1})\right]}{\rho_{N-1}}\\
&&-(E-e^{j\Delta x+1})=\frac{1}{\rho_{N-1}}\cdot\frac{\sigma_{N-1}^2\Delta t_{N-1}}{\Delta x^2}\left[(q_{N-1}e^{j\Delta x+c}-r_{N-1}E)\frac{\Delta x^2}{\sigma_{N-1}^2}+O(\Delta x^4)\right].
\end{eqnarray*} 
Note that $E>e^{j\Delta x+c}$ for $j\le k_1-1$. If $q_{N-1}\le r_{N-1}$ , then  $r_{N-1}E>q_{N-1}e^{j\Delta x+c}$   and therefore if $\Delta x$ is enough small, then we have $\psi_j<\varphi_j~(\forall j\le k_1-1)$ . Thus in our case, since $\exists j~(j\le k_1-1):~\psi_j>\varphi_j$, we must have $q_{N-1}>r_{N-1}$ . Then for sufficiently small $\Delta x$ , we have $j_{N-1}=\max\{j\le k_1-1:q_{N-1}e^{j\Delta x+c}\le r_{N-1}E\}$  and therefore $j_{N-1}\Delta x+c\le\ln\frac{r_{N-1}}{q_{N-1}}E$. If $j=j_{N-1}+1$ , then   $q_{N-1}e^{j\Delta x+c}>r_{N-1}E$ and thus $(j_{N-1}+1)\Delta x+c>\ln\frac{r_{N-1}}{q_{N-1}}E$ . So we have
$$ \ln\frac{r_{N-1}}{q_{N-1}}E-\Delta x<j_{N-1}\Delta x+c\le \ln\frac{r_{N-1}}{q_{N-1}}E.$$

Thus combining this inequality with \eqref{eq4-17}, we have the following theorem which provides an estimate of the approximated optimal exercise boundary near the maturity. 
\begin{theorem} \label{theorem4-4}
$\quad \ln\min\left(E,~\frac{r_{N-1}}{q_{N-1}}E\right)-2\Delta x\le j_{N-1}\Delta x+c\le \ln\min\left(E,~\frac{r_{N-1}}{q_{N-1}}E\right).$
\end{theorem} 

For fixed $\Delta x$ , the {\it approximated optimal exercise boundary} $x=\rho_{\Delta x}(t)$  is defined as follows:
$$\rho_{\Delta x}(t)=\frac{t-t_n}{t_{n+1}-t_n}(j_{n+1}\Delta x+c)+\frac{t_{n+1}-t}{t_{n+1}-t_n}(j_{n}\Delta x+c),~t\in [t_n,t_{n+1}],n=0,\cdots,N-2. $$
{\bf Corollary} 
(i) {\it $\rho_{\Delta x}(t_{N-1})\in\left[\ln\min\left(E,~\frac{r_{N-1}}{q_{N-1}}E\right)-2\Delta x,~ \ln\min\left(E,~\frac{r_{N-1}}{q_{N-1}}E\right)\right]$}.\\
(ii) $\rho_{\Delta x}(t)$ {\it is increasing on} $t$. 
 
%
%

\section{Convergence of the Explicit Difference Scheme and BTM for American Put Option.}

In this section we will prove that the explicit difference scheme \eqref{eq4-10} and \eqref{eq4-11} for American put option converges to the viscosity solution to the variational inequality \eqref{eq4-3} and using it prove the monotonic properties of the price of American put option and its optimal exercise boundary.  

We denote by $l^\infty(Z)$  the Banach space of all bounded two sided sequences with sup norm. In  $l^\infty(Z)$, we define $(U_j)\le (V_j)\Leftrightarrow U_j\le V_j,\forall j\in Z$. 

For fixed every $n$, the two sided sequence ${\bf U}^n=(\cdots,U_j,\cdots)^\infty_{j=-\infty}$  of American put option's prices $U_j,\forall j\in Z$  given by \eqref{eq4-10} and \eqref{eq4-11} is bounded from  theorem \ref{theorem4-1}. 

If we denote the right side of \eqref{eq4-11} by $({\bf F}_n{\bf U}^{n+1})_j$ , then ${\bf F}_n$ defines an operator 
\begin{equation} \label{eq5-1} 
{\bf U}^n:={\bf F}_n{\bf U}^{n+1}=\{({\bf F}_n{\bf U}^{n+1})_j\}_{j=-\infty}^\infty
\end{equation}
sending the sequence ${\bf U}^{n+1}$ of $t_{n+1}$ -time prices to the sequence ${\bf U}^{n}$ of $t_{n}$ -time prices. The operator ${\bf F}_n$ depends not only on $n$  and $\Delta x$  but also on $t_n$ and $\Delta t_n$.
\begin{lemma} \label{lemma5-1}
{\it If ~ $0<\alpha\le 1$, $\left|\frac{\Delta x}{\sigma^{2}_n}\left(r_n-q_n-\frac{\sigma^{2}_n}{2}\right)\right|<1$, then ${\bf F}_n$ is increasing,~that is,}  
$${\bf U}\le {\bf V},~{\bf U},{\bf V}\in l^\infty(Z) \Rightarrow {\bf F}_n{\bf U}\le {\bf F}_n{\bf V}.$$
\end{lemma} 

(Proof) Noting that from the assumption we have $1-\alpha\ge 0$ and $0<a_n<1$, the required result easily comes from \eqref{eq4-11}. (QED) 
\begin{lemma} \label{lemma5-2}
{\it If ${\bf U}\in l^\infty(Z)$ and ${\bf K}=(\cdots,K,K,K\cdots)$ with $K\ge0$, then }
$$ {\bf F}_n({\bf U}+{\bf K})\le {\bf F}_n{\bf U}+{\bf K}.$$
\end{lemma} 

(Proof) Since $\rho_n>1$ , we have 
\begin{eqnarray*}
&&{\bf F}_n({\bf U}+{\bf K})=\\
&&=\left(\max\left\{\frac{1}{\rho_n}\left[(1-\alpha)(U_j+K)+\alpha\left(a_n(U_{j+1}+K)+(1-a_n)(U_{j-1}+K)\right)\right],~\varphi_j\right\}\right)^\infty_{j=-\infty}\\
&&\le\left(\max\left\{\frac{1}{\rho_n}\left[(1-\alpha)U_j+\alpha\left(a_nU_{j+1}+(1-a_n)U_{j-1}\right)\right],~\varphi_j\right\}+\max\left\{\frac{K}{\rho_n},~0\right\}\right)^\infty_{j=-\infty}\\
&&\le{\bf F}_n{\bf U}+{\bf K}.
\end{eqnarray*}
(QED) \\

Define the extension function $u_{\Delta x}(x,~t)$  as follows: 
\begin{eqnarray} \label{eq5-2} 
&&x\in [(j-1/2)\Delta x+c,~(j+1/2)\Delta x+c), t\in[t_n,t_{n+1}) \Rightarrow u_{\Delta x}(x,~t):=U_j^n,\\
&&x\in [(j-1/2)\Delta x+c,~(j+1/2)\Delta x+c), t\in[t_N,T) ~~\Rightarrow u_{\Delta x}(x,~t):=U_j^N.   \label{eq5-3}
\end{eqnarray}

{\bf Remark 5.1}. Here $u_{\Delta x}$ is piecewise continuous function. The following discussion is also true if we define  $u_{\Delta x}$ as a continuous function interpolating the data set $(j\Delta x+c,t_n;U_j^n)$ .  

From Theorem \ref{theorem4-1}, we have
\begin{eqnarray} \label{eq5-4} 
0\le u_{\Delta x}(x,~t)\le E.
\end{eqnarray}
Therefore, for every fixed $t$, if we denote ${\bf u}_{\Delta x}(\bullet,~t):=\{u_{\Delta x}(x,~t):x\in {\bf R}\}$, then we have
$${\bf u}_{\Delta x}(\bullet,~t)\in l^\infty(Z).$$

When $t\in[t_n,t_{n+1}),~n=0,\cdots,N-1$, if we define 
\begin{eqnarray} \label{eq5-5} 
\Delta t=\Delta t(t,\Delta x):=\frac{t-t_n}{t_{n+1}-t_n}\Delta t_{n+1}+\frac{t_{n+1}-t}{t_{n+1}-t_n}\Delta t_{n},
\end{eqnarray}
then  $t+\Delta t\in[t_{n+1},t_{n+2})$ and thus we have
\begin{eqnarray} \label{eq5-6} 
{\bf u}_{\Delta x}(\bullet,~t)={\bf F}_n{\bf u}_{\Delta x}(\bullet,~t+\Delta t).
\end{eqnarray}
$\frac{\Delta t}{\Delta t_n}=\frac{t-t_n}{t_{n+1}-t_n}\cdot\frac{\sigma^2_{n+1}}{\sigma^2_n}+\frac{t_{n+1}-t}{t_{n+1}-t_n}\cdot 1$ is a convex linear combination of $\frac{\sigma^2_{n+1}}{\sigma^2_n}$ and $1$. Thus if $\sigma(t)$ is continuous, then this ratio converges to $1$ when $\Delta x \rightarrow 0$ . Like this, we proved the following lemma: 
\begin{lemma} \label{lemma5-3}
{\it Assume that  $\sigma(t)$ is continuous. Then }
$\lim_{\Delta x \rightarrow 0} \frac{\Delta t(t,\Delta x)}{\Delta t_n}=1.$
\end{lemma} 

In order to prove the convergence, we recall the notion of viscosity solutions.\\

Let $USC([0, T]\times R)~ (LSC([0, T]\times R))$ be the space of all upper (lower) semi-continuous functions defined on $[0, T]\times R$. If $u\in USC([0, T]\times R)~ (LSC([0, T]\times R))$ satisfies the following two conditions, then $u$ is called the {\it viscosity subsolution (supersolution)} of the variational inequality \eqref{eq4-3}: 

(i) $u(x,~T)\le (\ge) \varphi(x)$,

(ii)  If $\Phi\in C^{1,2}([0, T]\times R)$ and $u-\Phi$ attains its local maximum (minimum) at $(x, t)\in [0, T]\times R$, we have 
$$\min\left\{-\frac{\partial \Phi}{\partial t}-\frac{\sigma(t)^{2}}{2}\frac{\partial^2 \Phi}{\partial x^2}-\left(r(t)-q(t)-\frac{\sigma(t)^{2}}{2}\right)\frac{\partial \Phi}{\partial x}+r(t)u,u-\varphi\right\}_{(x,t)}\le (\ge) 0$$

$u\in C([0, T]\times R)$ is called the {\it viscosity solution} of the variational inequality \eqref{eq4-3} if it is both viscosity subsolution and viscosity supersolution of \eqref{eq4-3}.
\begin{lemma} \label{lemma5-4}
(Comparison lemma)\cite{CIL} {\it If $u$ and $v$ are viscosity subsolution and supersolution of \eqref{eq4-3} and $|u(x,t)|,|v(x,t)|\le E$, then $u \le v$. }
\end{lemma} 
\begin{theorem}\cite{CIL} \label{theorem5-1}
If $r(t),~q(t)$ are continuous on $[0,T]$, then the problem \eqref{eq4-3} has a unique viscosity solution. Furtheremore, there exists an optimal exercise boundary $\rho(t):[0,T]\rightarrow R$ (continuous function) such that if $x<\rho(t)$, then $u(x,t)=\varphi(x)$ ; if $x>\rho(t)$, then  $u(x,t)>\varphi(x)$ and in this region $u(x,t)$ is a classical solution to the equation 
$$-\frac{\partial u}{\partial t}-\frac{\sigma(t)^{2}}{2}\frac{\partial^2 u}{\partial x^2}-\left(r(t)-q(t)-\frac{\sigma(t)^{2}}{2}\right)\frac{\partial u}{\partial x}+r(t)u=0.$$
\end{theorem} 

{\bf Remark 5.2}. It is easy to show $\rho(T)=\ln E \min [1,r(T)/q(T)]$, using the way of \cite{jia}.
\begin{theorem} \label{theorem5-2}
Suppose that $u(x,t)$ is the viscosity solution of \eqref{eq4-3}. Assume that $0<\alpha\le 1$, $r(t)/\sigma^2(t)$ is increasing and $q(t)/\sigma^2(t)$ decreasing on $t$. Then we have

(i) $u_{\Delta x}(x,t)$   converges to $u(x,t)$ as $\Delta x\rightarrow 0$.

(ii) $\rho_{\Delta x}(t)$ converges to $\rho(t)$ as $\Delta x\rightarrow 0$.
\end{theorem} 
\begin{proof}
Suppose that $u(x,t)$ is the viscosity solution of  \eqref{eq4-3} and denote
\begin{eqnarray} \label{eq5-7} 
&&u^*(x,t)=\lim_{\Delta x\rightarrow 0,~(y,s)\rightarrow (x,t)} \sup u_{\Delta x}(y,s),\nonumber\\
&&u_*(x,t)=\lim_{\Delta x\rightarrow 0,~(y,s)\rightarrow (x,t)} \inf u_{\Delta x}(y,s)
\end{eqnarray}
From \eqref{eq5-4}, $u^*$ and $u_*$ are well defined and we have $0\le u_*(x,t)\le u^*(x,t)\le E$ . Obviously $u^*\in USC([0, T]\times R)$ and $u_*\in LSC([0, T]\times R)$. If we prove that $u^*$ and $u_*$ are subsolution and supersolution of \eqref{eq4-3}, respectively, then from lemma \ref{lemma5-4}, we have $u^*\le u_*$  and thus $u^*=u_*=u(x,t)$  becomes a viscosity solution of \eqref{eq4-3}, and therefore we have the convergence of the approximated solution $u_{\Delta x}(x,t)$ .

We will prove that $u^*$ is a subsolution of \eqref{eq4-3}. (The fact that  $u_*$ is a supersolution is similarly proved.)  From the definition \eqref{eq5-3}, we can easily know that 
$$u^*(x,T)=\varphi(x)=(E-e^x)^+.$$
Suppose that $\phi\in C^{1,2}([0, T]\times R)$ and $u^*-\phi$ attains a local maximum at $(x_0,t_0)\in [0,T)\times R$. We might as well assume that $(u^*-\phi)(x_0,t_0)=0$ and $(x_0,t_0)$ is a strict local maximum on $B_r=\{(x,t):t_0\le t\le t_0+r,~|x-x_0|\le r\}, r>0$ . Let $\Phi = \phi-\epsilon, \epsilon>0$, then $u^*-\Phi$ attains a strict local maximum at $(x_0,t_0)$ and 
\begin{eqnarray} \label{eq5-8} 
(u^*-\Phi)(x_0,t_0)>0.
\end{eqnarray}
From the definition of $u^*$, there exists a sequence $u_{\Delta x_k}(y_k,s_k)$  such that $\Delta x_k\rightarrow 0, y_k\rightarrow x_0, s_k\rightarrow t_0$ and 
\begin{eqnarray} \label{eq5-9} 
\lim_{k\rightarrow\infty} u_{\Delta x_k}(y_k,s_k)=u^*(x_0,t_0).
\end{eqnarray}
If we denote the global maximum point of $u_{\Delta x_k}-\Phi$ on $B_r$ by $(\hat{y}_k, \hat{S}_k)$, then there exists a subsequence $u_{\Delta x_{k_i}}(\hat{y}_k, \hat{S}_k)$  such that
\begin{eqnarray} \label{eq5-10} 
\Delta x_{k_i}\rightarrow 0, (\hat{y}_{k_i}, \hat{S}_{k_i})\rightarrow (x_0,t_0), (u_{\Delta x_{k_i}}-\Phi)(\hat{y}_{k_i}, \hat{S}_{k_i})\rightarrow (u^*-\Phi)(x_0,t_0)~\text{as}~ k_i\rightarrow\infty.
\end{eqnarray}
Indeed, suppose $(\hat{y}_{k_i},  \hat{S}_{k_i})\rightarrow (\hat{y},\hat{s})$, then from \eqref{eq5-9} we have
$$(u^*-\Phi)(x_0,t_0)=\lim_{k_i\rightarrow\infty} (u_{\Delta x_{k_i}}-\Phi)(y_{k_i},s_{k_i})\le \lim_{k_i\rightarrow\infty} (u_{\Delta x_{k_i}}-\Phi)(\hat{y}_{k_i},  \hat{S}_{k_i})\le (u^*-\Phi)(\hat{y},\hat{s}).$$
Therefore we have $(\hat{y},\hat{s})=(x_0,t_0)$, since $(x_0,t_0)$ is a strict local maximum of $(u^*-\Phi)$. Thus for sufficiently large $k_i$, $\Delta t=\Delta t(\Delta x_{k_i})$ defined by \eqref{eq5-5} is  small enough and if $(x,\hat{S}_{k_i}+\Delta t(\Delta x_{k_i}))\in B_r$, then we have
$$(u_{\Delta x_{k_i}}-\Phi)(x,  \hat{S}_{k_i}+\Delta t(\Delta x_{k_i}))\le (u_{\Delta x_{k_i}}-\Phi)(\hat{y}_{k_i},  \hat{S}_{k_i}),$$
that is ,
\begin{eqnarray} \label{eq5-11} 
u_{\Delta x_{k_i}}(x,  \hat{S}_{k_i}+\Delta t(\Delta x_{k_i}))\le \Phi(x,  \hat{S}_{k_i}+\Delta t(\Delta x_{k_i}))+(u_{\Delta x_{k_i}}-\Phi)(\hat{y}_{k_i},  \hat{S}_{k_i}).
\end{eqnarray}
From \eqref{eq5-8} and \eqref{eq5-10}, we have
\begin{eqnarray} \label{eq5-12} 
(u_{\Delta x_{k_i}}-\Phi)(\hat{y}_{k_i},  \hat{S}_{k_i})>0~(\text{for sufficiently large}~k_i).
\end{eqnarray}
For every $k_i$, define $t_n$ and $\Delta t_n=\frac{\alpha\Delta x^2_{k_i}}{\sigma^2(t_n)}$ as in \eqref{eq4-4} with $\Delta x_{k_i}$. Select $t_{n_i}$ and $j_{k_i}=j$ such that $\hat{y}_{k_i}\in [(j-1/2)\Delta x_{k_i}+c,(j+1/2)\Delta x_{k_i}+c)$ , $\hat{S}_{k_i}\in [t_{n_i}, t_{n_i+1})$, and simply denote $t_{n_i}=t_n$ and $j=j_{k_i}$. Then from \eqref{eq5-6}, lemma \ref{lemma5-1} with \eqref{eq5-11} and lemma \ref{lemma5-2} we have
\begin{eqnarray*} 
u_{\Delta x_{k_i}}(\hat{y}_{k_i},  \hat{S}_{k_i})&=&U_j^n=({\bf F}_n{\bf U}^{n+1})_j=[{\bf F}_n{\bf u}_{\Delta t_{k_i}}(\bullet,\hat{S}_{k_i}+\Delta t(\Delta x_{k_i}))](\hat{y}_{k_i})\\
&\le &\{{\bf F}_n[{\bf\Phi}(\bullet,\hat{S}_{k_i}+\Delta t)+(u_{\Delta x_{k_i}}-\Phi)(\hat{y}_{k_i},  \hat{S}_{k_i})]\}(\hat{y}_{k_i})\\
&\le &\{{\bf F}_n[{\bf\Phi}(\bullet,\hat{S}_{k_i}+\Delta t)]\}(\hat{y}_{k_i})+(u_{\Delta x_{k_i}}-\Phi)(\hat{y}_{k_i},  \hat{S}_{k_i}).
\end{eqnarray*}
Thus we have
\begin{eqnarray*} 
\Phi(\hat{y}_{k_i},  \hat{S}_{k_i})-\{{\bf F}_n[{\bf\Phi}(\bullet,\hat{S}_{k_i}+\Delta t)]\}(\hat{y}_{k_i})\le 0.
\end{eqnarray*}
Therefore using \eqref{eq4-11} we have
\begin{eqnarray*} 
&&\Phi(\hat{y}_{k_i},  \hat{S}_{k_i})-\max\left\{\frac{1}{1+r_n\Delta t_n}\left[\left(1-\frac{\sigma^2_n\Delta t_n}{\Delta x_{k_i}}\right)\Phi(\hat{y}_{k_i}, \hat{S}_{k_i}+\Delta t(\Delta x_{k_i}))+\right.\right.\\
&&\left.\left.+\frac{\sigma^2_n\Delta t_n}{\Delta x_{k_i}}[a_n\Phi(\hat{y}_{k_i}+\Delta x_{k_i}, \hat{S}_{k_i}+\Delta t)+(1-a_n)\Phi(\hat{y}_{k_i}-\Delta x_{k_i}, \hat{S}_{k_i}+\Delta t)]\right],\varphi_j\right\}\le 0.
\end{eqnarray*}
This inequality is equivalent to the following.
\begin{eqnarray*} 
&&\min\left\{\frac{\Delta t_n}{1+r_n\Delta t_n}\left[\frac{\Phi(\hat{y}_{k_i},  \hat{S}_{k_i})-\Phi(\hat{y}_{k_i}, \hat{S}_{k_i}+\Delta t(\Delta x_{k_i}))}{\Delta t_n}- \right.\right.\\
&&\quad\quad-\frac{\sigma^2_n}{2}\frac{\Phi(\hat{y}_{k_i}+\Delta x_{k_i}, \hat{S}_{k_i}+\Delta t)-2\Phi(\hat{y}_{k_i}, \hat{S}_{k_i}+\Delta t)+\Phi(\hat{y}_{k_i}-\Delta x_{k_i}, \hat{S}_{k_i}+\Delta t)}{2\Delta x_{k_i}}-\\
&&\quad\quad-\left(r_n-q_n-\frac{\sigma^2_n}{2}\right)\frac{\Phi(\hat{y}_{k_i}+\Delta x_{k_i}, \hat{S}_{k_i}+\Delta t)-\Phi(\hat{y}_{k_i}-\Delta x_{k_i}, \hat{S}_{k_i}+\Delta t)}{2\Delta x_{k_i}}+\\
&&\quad\quad\left.\left.+r_n\Phi(\hat{y}_{k_i}, \hat{S}_{k_i})\right],\Phi(\hat{y}_{k_i}, \hat{S}_{k_i})-\varphi_j\right\}\le 0.
\end{eqnarray*}
Noting that $\frac{\Delta t_n}{1+r_n\Delta t_n}>0$, we have 
\begin{eqnarray*} 
&&\min\left\{\frac{\Phi(\hat{y}_{k_i},  \hat{S}_{k_i})-\Phi(\hat{y}_{k_i}, \hat{S}_{k_i}+\Delta t(\Delta x_{k_i}))}{\Delta t_n}- \right.\\
&&\quad\quad-\frac{\sigma^2_n}{2}\frac{\Phi(\hat{y}_{k_i}+\Delta x_{k_i}, \hat{S}_{k_i}+\Delta t)-2\Phi(\hat{y}_{k_i}, \hat{S}_{k_i}+\Delta t)+\Phi(\hat{y}_{k_i}-\Delta x_{k_i}, \hat{S}_{k_i}+\Delta t)}{2\Delta x_{k_i}}-\\
&&\quad\quad-\left(r_n-q_n-\frac{\sigma^2_n}{2}\right)\frac{\Phi(\hat{y}_{k_i}+\Delta x_{k_i}, \hat{S}_{k_i}+\Delta t)-\Phi(\hat{y}_{k_i}-\Delta x_{k_i}, \hat{S}_{k_i}+\Delta t)}{2\Delta x_{k_i}}+\\
&&\quad\quad\left.+r_n\Phi(\hat{y}_{k_i}, \hat{S}_{k_i}),\Phi(\hat{y}_{k_i}, \hat{S}_{k_i})-\varphi_j\right\}\le 0.
\end{eqnarray*}
This inequality is equivalent to the following.
\begin{eqnarray*} 
&&\min\left\{\left[\frac{\Phi(\hat{y}_{k_i},  \hat{S}_{k_i})-\Phi(\hat{y}_{k_i}, \hat{S}_{k_i}+\Delta t(\Delta x_{k_i}))}{\Delta t(\Delta x_{k_i})}\right]\frac{\Delta t(\Delta x_{k_i})}{\Delta t_n}- \right.\\
&&\quad\quad-\frac{\sigma^2_n}{2}\frac{\Phi(\hat{y}_{k_i}+\Delta x_{k_i}, \hat{S}_{k_i}+\Delta t)-2\Phi(\hat{y}_{k_i}, \hat{S}_{k_i}+\Delta t)+\Phi(\hat{y}_{k_i}-\Delta x_{k_i}, \hat{S}_{k_i}+\Delta t)}{2\Delta x_{k_i}}-\\
&&\quad\quad-\left(r_n-q_n-\frac{\sigma^2_n}{2}\right)\frac{\Phi(\hat{y}_{k_i}+\Delta x_{k_i}, \hat{S}_{k_i}+\Delta t)-\Phi(\hat{y}_{k_i}-\Delta x_{k_i}, \hat{S}_{k_i}+\Delta t)}{2\Delta x_{k_i}}+\\
&&\quad\quad\left.+r_n\Phi(\hat{y}_{k_i}, \hat{S}_{k_i}),\Phi(\hat{y}_{k_i}, \hat{S}_{k_i})-\varphi_j\right\}\le 0.
\end{eqnarray*}
Let $k_i\rightarrow\infty$, then $\Delta x_{k_i}\rightarrow 0$. From lemma \ref{lemma5-3}, we have $\frac{\Delta t(\Delta x_{k_i})}{\Delta t_n}\rightarrow 1$. Thus we have
\begin{eqnarray*} 
\min\left\{-\frac{\partial \Phi}{\partial t}-\frac{\sigma^2(t)}{2}\frac{\partial^2\Phi}{\partial x^2}-\left(r(t)-q(t)-\frac{\sigma^2(t)}{2}\right)\frac{\partial\Phi}{\partial x}+r(t)\Phi,~\Phi-\varphi\right\}_{(x_0,~t_0)}\le 0.
\end{eqnarray*}
(Here we considered $(\hat{y}_{k_i},~\hat{S}_{k_i})\rightarrow (x_0,~t_0)$ and $\varphi_{j_{k_i}}\rightarrow \varphi(x_0)$. Let $\epsilon\rightarrow 0$ , then we have
\begin{eqnarray*} 
\min\left\{-\frac{\partial \phi}{\partial t}-\frac{\sigma^2(t)}{2}\frac{\partial^2\phi}{\partial x^2}-\left(r(t)-q(t)-\frac{\sigma^2(t)}{2}\right)\frac{\partial\phi}{\partial x}+r(t)\phi,~\phi-\varphi\right\}_{(x_0,~t_0)}\le 0.
\end{eqnarray*}
Since $u^*(x_0,~t_0) = \phi(x_0,~t_0)$,  $u^*$ is a subsolution of \eqref{eq4-3}. Thus we proved (i).

Now we will prove $\rho_{\Delta x}(t)$ converges to $\rho(t)$ as $\Delta x\rightarrow 0$. The main idea is from \cite{JD2}. First,  from Corollary of theorem \ref{theorem4-4} and Remark 9, we have
$$\lim_{\Delta x\rightarrow 0}\rho_{\Delta x}(t_{N-1})=\ln\min\left[E,\frac{r(T)}{q(T)}E\right]=\rho(T).$$
Now fix $t_0\in [0,T)$ and suppose $x<\lim_{\Delta x\rightarrow 0}\sup\rho_{\Delta x}(t_0)$. Then there exists a sequence $\Delta x_k$ such that $\Delta x_k\rightarrow 0$ and $\lim_{k\rightarrow\infty}\rho_{\Delta x_k}(t_0)>x$. Denote by $\{t_n^{(k)}\}$ the time partition corresponding to $\Delta x_k$ and let $t_0\in [t_{n_0-1}^{(k)},t_{n_0}^{(k)})$. Then $t_{n_0}^{(k)}\rightarrow t_0$ as $k\rightarrow\infty$. Since $\rho_{\Delta x}$ is increasing, we have 
$$\rho_{\Delta x_k}(t_{n_0}^{(k)})\ge\rho_{\Delta x_k}(t_0).$$
Select $x_k$ such that $\lim \Delta x_k=x$. For sufficiently large $k$, 
$x_k<\rho_{\Delta x_k}(t_{n_0}^{(k)})=j_{n_0}\Delta x_k+c$ 
and thus we have $u_{\Delta x_k}(x_k,t_{n_0}^{(k)})=\varphi(x_k)$ (since $x_k$ is in the exercise region). Thus we have $\Delta x_k\rightarrow 0\Rightarrow u(x,t_0)=\varphi(x)\Rightarrow x\le \rho(t_0)$, so we have 
$$\lim\sup \rho_{\Delta x}(t_0)\le\rho(t_0).$$ 
Now we will prove $\lim\inf\rho_{\Delta x}(t_0)\ge\rho(t_0)$. Assume that there exists $\epsilon>0$ such that $\lim\inf\rho_{\Delta x}(t_0)<\rho(t_0)-2\epsilon$ . From the fact that $\rho(t)$ is continuous, there exists $\delta >0$ such that $\lim\inf\rho_{\Delta x}(t_0)<\rho(t)-2\epsilon, t\in[t_0-\delta,t_0+\delta]$. Therefore there exists a sequence $\Delta x_k\rightarrow 0$ such that $\rho_{\Delta x_k}(t_0)<\rho(t)-2\epsilon, t\in[t_0-\delta,t_0+\delta]$. Now let $\underline{\rho}:=\min\{\rho(t):t\in [t_0-\delta,t_0+\delta]\}$ and $Q=\{(x,t):t\in(t_0-\delta,t_0], x\in (\underline{\rho}-2\epsilon,\underline{\rho}-\epsilon)\}$. Then since $\rho_{\Delta x_k}(t)$ is increasing, we have $\rho_{\Delta x_k}(t)\le \rho_{\Delta x_k}(t_0)\le \underline{\rho}-2\epsilon<x<\underline{\rho}-\epsilon$ for $(x,t)\in Q$. Therefore we have 
\begin{eqnarray} \label{eq5-13}          
\rho_{\Delta x_k}(t)<x<\rho(t)-\epsilon, ~(x,t)\in Q.
\end{eqnarray} 
From $\rho_{\Delta x_k}(t)<x$, let $u_{\Delta x_k}(x,t)=U_j^n$, then we have 
\begin{eqnarray*}
&&-\frac{U_j^{n+1}-U_j^{n}}{\Delta t}-\frac{\sigma^2(t_n)}{2}\frac{U_{j+1}^{n+1}-2U_j^{n+1}+U_{j-1}^{n+1}}{\Delta x^2}\\
&&-\left[r(t_n)-q(t_n)-\frac{\sigma^2(t_n)}{2}\right]\frac{U_{j+1}^{n+1}-U_{j-1}^{n+1}}{2\Delta x}+r(t_n)U_j^n=0.
\end{eqnarray*}
Letting $\Delta x_k\rightarrow 0$ , then $u_{\Delta x_k}(x,t)~(=U_j^n)$ converges to the viscosity solution $u(x,t)$ and from the regularity of viscosity solution (Theorem \ref{theorem5-1}), we have 
\begin{eqnarray} \label{eq5-14}             
\frac{\partial u}{\partial t}+\frac{\sigma(t)^{2}}{2}\frac{\partial^2 u}{\partial x^2}+\left(r(t)-q(t)-\frac{\sigma(t)^{2}}{2}\right)\frac{\partial u}{\partial x}-r(t)u=0,~(x,t)\in Q.
\end{eqnarray} 
On the other hand, from \eqref{eq5-13}, we have $x<\rho(t)-\epsilon$ , thus $x$ is in the exercise region of \eqref{eq4-3} and $u(x,t)=E-e^x$. So we have
$$\frac{\partial u}{\partial t}+\frac{\sigma(t)^{2}}{2}\frac{\partial^2 u}{\partial x^2}+\left(r(t)-q(t)-\frac{\sigma(t)^{2}}{2}\right)\frac{\partial u}{\partial x}-r(t)u=q(t)e^x-r(t)E,~(x,t)\in Q.$$
This contradicts to \eqref{eq5-14}. Thus we proved   
$$\lim\inf\rho_{\Delta x}(t_0)\ge\rho(t_0)\ge\lim\sup\rho_{\Delta x}(t_0).$$ 
So we have $~\lim\rho_{\Delta x}(t_0)=\rho(t_0)$.(QED) 
\end{proof}\\

{\bf Corollary} (Monotonicity of American put option's price and optimal exercise boundary)  {\it Suppose that $u(x,t)$ is the viscosity solution of \eqref{eq4-3}. Assume that   $r(t)/\sigma^2(t)$ is increasing and $q(t)/\sigma^2(t)$  is decreasing on $t$. Then we have}

(i) {\it $u(x,t)$ is decreasing on $x$ and $t$.}

(ii) {\it $\rho(t)$ is increasing on $t$.}
\\

(Proof) (i) comes from Theorem \ref{theorem4-1} (ii), Theorem \ref{theorem4-2} and Theorem \ref{theorem5-2} (i).

(ii) comes from (ii) of Corollary of Theorem \ref{theorem4-4} and Theorem \ref{theorem5-2} (ii). (QED)\\

\begin{lemma} \cite{JD2} \label{lemma5-5}
{\it Let $\Omega\subset R^m$ and $f_n(x_1,\cdots,x_m)$ be pointwise convergent to a continuous function $f(x_1,\cdots,x_m)$. Assume that $f_n$ and $f$ are monotone on $\Omega$. Then $f_n$ uniformly converges to $f$ on any compact subset of  $\Omega$}.
\end{lemma}

\begin{theorem}  \label{theorem5-3}
When $\Delta x\rightarrow 0$, then $u_{\Delta x}(x,t)$ uniformly converges to $u(x,t)$ on any bounded subdomain of  $[0,T]\times R$ and $\rho_{\Delta x}(t)$ uniformly converges to $\rho(t)$.
\end{theorem}
(Proof) From the result of Theorem \ref{theorem5-3}, $u(x,t)$ and $\rho(t)$ are both monotone. Thus from lemma \ref{lemma5-5}, we have the desired results. (QED)\\

{\bf Remark 5.3}. Consider the convergence of the binomial tree methods. As shown in section 3, in the lattice $Q_c=\{(x_j,~t_n):x_j=j\Delta x+c,0\le n\le N-1,j\in Z\}$, the explicit difference scheme \eqref{eq4-12} with $\alpha=\sigma^2_n\Delta t_n/\Delta x^2=1$ coincides with BTM in the sense of neglecting $O(\Delta x^3)$.

Let  $U_j^n$ be the prices by the explicit difference scheme \eqref{eq4-12} and $V_j^n$ the BTM prices. Then we have
\begin{eqnarray*}
&&U_j^n=\max\left\{\frac{1}{\rho_n}\left[a_n U_{j+1}^{n+1}+(1-a_n)U_{j-1}^{n+1}\right],~(E-S_0e^{j\Delta x})^+\right\}, \\
&&V_j^n=\max\left\{\frac{1}{\rho_n}\left[\theta_nV_{j+1}^{n+1}+(1-\theta_n)V_{j-1}^{n+1}\right], (E-S_0e^{j\Delta x})^+\right\} .
\end{eqnarray*}
Let  $u=e^{\Delta x}$. Then we have 
\begin{eqnarray*}
&&|V_j^n-U_j^n|\le\frac{1}{\rho_n}\left[|\theta_nV_{j+1}^{n+1}-a_n U_{j+1}^{n+1}|+|(1-\theta_n)V_{j-1}^{n+1}-(1-a_n)U_{j-1}^{n+1}|\right]\\
&&~~~\le\frac{1}{\rho_n}\left[\theta_n|V_{j+1}^{n+1}-U_{j+1}^{n+1}|+|\theta_n-a_n| U_{j+1}^{n+1}+(1-\theta_n)|V_{j-1}^{n+1}-U_{j-1}^{n+1}|+|\theta_n-a_n)|U_{j-1}^{n+1}\right]\\
&&~~~\le\frac{1}{\rho_n}\left[\|{\bf V}^{n+1}-{\bf U}^{n+1}\|_{l^\infty(Z)}+2O(\Delta x^3)\|{\bf U}^{n+1}\|_{l^\infty(Z)}\right].
\end{eqnarray*}
Here we considered lemma \ref{lemma4-1}: $\theta_n=a_n+O(\Delta x^3)$. Note that $\|{\bf U}^{n+1}\|_{l^\infty(Z)}\le E$, then we have
$$|V_j^n-U_j^n|\le\frac{1}{\rho_n}\left[\|{\bf V}^{n+1}-{\bf U}^{n+1}\|_{l^\infty(Z)}+2E\cdot O(\Delta x^3)\right].$$
Since $r_n/\sigma^2_n$ is increasing, we have 
$$\exp(-r_n\Delta t_n)=\exp(-r_n\Delta x^2/\sigma^2_n)\le\exp(-r_0\Delta x^2/\sigma^2_0)=:A,$$
and using $\rho_n^{-1}=(1+r_n\Delta t_n)=\exp(-r_n\Delta t_n)+O(\Delta t^2_n)$ , then we  have
$$|V_j^n-U_j^n|\le A\|{\bf V}^{n+1}-{\bf U}^{n+1}\|_{l^\infty(Z)}+2AE\cdot O(\Delta x^3).$$
Therefore  we have
\begin{eqnarray*}
\|{\bf V}^{n}-{\bf U}^{n}\|_{l^\infty(Z)}&\le&A\|{\bf V}^{n+1}-{\bf U}^{n+1}\|_{l^\infty(Z)}+A2E\cdot O(\Delta x^3)\\
&\le&A^2\|{\bf V}^{n+2}-{\bf U}^{n+2}\|_{l^\infty(Z)}+(A+A^2)2E\cdot O(\Delta x^3)\\
&\cdots&\cdots\cdots\\
&\le&A^{N-n}\|{\bf V}^{N}-{\bf U}^{N}\|_{l^\infty(Z)}+(A+A^2+\cdots +A^{N-n})2E\cdot O(\Delta x^3).
\end{eqnarray*}
Here noting that $\|{\bf V}^{N}-{\bf U}^{N}\|_{l^\infty(Z)}=\max\{|V_j^N-U_j^N|\}=0$,  we have
$$ \|{\bf V}^{n}-{\bf U}^{n}\|_{l^\infty(Z)}\le\frac{A}{1-A}2E\cdot O(\Delta x^3)=O(\Delta x).$$
Thus from theorem \ref{theorem5-2}, when $\Delta x\rightarrow 0$, the BTM prices $V_j^n$  converges to the viscosity solution to the variational inequality \eqref{eq4-3}.

\section{Call-put parity in BTM for American option and it’s applications}

\indent

In order to prove the monotony of prices by American call option's price on $t$, we need not only Lemma \ref{lemma3-1} and  Lemma \ref{lemma3-2}  but also the following lemmas which are proved in the same way as \cite{jia}.

\begin{lemma}\label{lemma6-1}
If $\theta_n=\frac{\rho_n/\eta_n-d}{u-d},~\theta_n^\prime=\frac{\eta_n/\rho_n-d}{u-d}$
\begin{equation}\label{eq6-1}
\frac{\theta_n u}{\rho_n}=\frac{1-\theta_n^\prime}{\eta_n},~~\frac{(1-\theta_n)d}{\rho_n}=\frac{\theta_n^\prime}{\eta_n}
\end{equation}
\end{lemma}
\begin{proof}
It is proved in the same way as \cite{jia}. (Q.E.D)
\end{proof}
\begin{lemma}[Homogeneity]\label{lemma6-2}
If we denote BTM price of American put option by $P(S_j;E)$ and American call option by $C(S_j;E)$($E$ is the exercise price), then we have
\begin{equation}\label{eq6-2}
C(\alpha S_j;\alpha E)=\alpha C(S_j;E),~~P(\alpha S_j;\alpha E)=\alpha P(S_j;E)
\end{equation}
\end{lemma}

\begin{proof}
It is proved in the same way as \cite{jia}   (Q.E.D)
\end{proof}

\indent

By using these two lemmas, we obtain the following call-put symmetry.

\begin{theorem}[Call-Put Symmetry in BTM]\label{theorem6-1}
Denote prices of American options with the underlying asset's price $S_j$, the exercise price $E$, the interest rate $r_n$ and dividend rate $q_n$ by
$$C(S_j,E,n)=C(S_j,E,\rho_n,\eta_n)~and~P(S_j,E,n)=P(S_j,E,\rho_n,\eta_n),$$
respectively. The we have
\begin{equation}\label{eq6-3}
C(S_j,E,\rho_n,\eta_n)=P(E,S_j,\eta_n,\rho_n).
\end{equation}
$(\rho_n=1+r_n \Delta t_n,~\eta_n=1+q_n \Delta t_n)$
\end{theorem}

\begin{proof}
In the case of $n=N,~C(S_j,E,N)=(S_j-E)+=P(E,S_j,N)$ and we have \eqref{eq6-3}. Now we assume that \eqref{eq6-3} holds for $n=k+1$ and prove when $n=k$. From \eqref{eq2-4}, the BTM price of American call option is
\begin{eqnarray}
C(S_j,E,\rho_n,\eta_n)&=&max \left\{\frac{1}{\rho_n}[\theta_n C(S_{j+1},E,n+1)+(1-\theta_n)C(S_{j-1},E,n+1)],(S_j-E)^+ \right\}\nonumber
\\
&=&max \left\{\frac{1}{\rho_n}[\theta_n C(S_j u,E,n+1)+(1-\theta_n)C(S_j d,E,n+1)],(S_j-E)^+ \right\}\nonumber
\end{eqnarray}
By using Lemma \ref{lemma6-1} and Lemma \ref{lemma6-2}(homogeneity), we have
\begin{eqnarray}
C(S_j,E,\rho_n,\eta_n)&=&max \left\{\frac{1}{\rho_n}[\theta_n u C(S_j,Ed,n+1)+(1-\theta_n)d C(S_j,Eu,n+1)],(S_j-E)^+ \right\}\nonumber
\\
&=&max \left\{\frac{1}{\eta_n}[(1-\theta^\prime_n) C(S_j,Ed,n+1)+\theta^\prime_nC(S_j,Eu,n+1)],(S_j-E)^+ \right\}\nonumber
\end{eqnarray}
Using the assumption of induction and BTM price formula of put option, we have
\begin{eqnarray}
C(S_j,E,\rho_n,\eta_n)&=&max \left\{\frac{1}{\eta_n}[\theta^\prime_n C(S_j,Eu,n+1)+(1-\theta^\prime_n)C(S_j,Ed,n+1)],(S_j-E)^+ \right\}\nonumber
\\
&=&max \left\{\frac{1}{\eta_n}[\theta^\prime_n P(Eu,S_j,n+1)+(1-\theta^\prime_n)P(Ed,S_j,n+1)],(S_j-E)^+ \right\}\nonumber
\\
&=&P(E,S_j,\eta_n,\rho_n)\nonumber
\end{eqnarray}
We can prove the decreasing property of call option's price on $t$ by using these results. (Q.E.D)
\end{proof}

\begin{theorem}\label{theorem6-2}
$V_j^n~(n=0,1,\cdots,N,~j=0,\pm 1,\pm 2,\cdots)$ is the BTM price of American call option. Assume that $r(t)/\sigma^2(t)$ is decreasing on $t$ and $q(t)/\sigma^2(t)$ is increasing on $t$. Then we have
\begin{equation}\label{eq6-4}
V_j^{n-1}\le V_j^n.
\end{equation}
\end{theorem}

\begin{proof}
We prove by induction. From the property of American option, we have
$$
V_j^{N-1}\ge\varphi_j=(S_j-E)^+=V_j^N~~(j=0,\pm 1,\cdots)
$$
Therefore the assertion holds true for $n=N$. Inductively, we assume that $V_j^k\ge V_j^{k+1}.$
Then
$$
V_j^{k-1}=max\left\{\frac{1}{\rho_{k-1}}[\theta_{k-1}V_{j+1}^k+(1-\theta_{k-1})V_{j-1}^k],\varphi_j\right\}
$$
\begin{equation}\label{eq6-5}
=max\left\{\frac{1}{\rho_{k-1}}[\theta_{k-1}C(S_j u,E,k)+(1-\theta_{k-1})C(S_jd,E,k)],\varphi_j\right\}
\end{equation}
$$
\ge max\left\{\frac{1}{\rho_{k-1}}[\theta_{k-1}C(S_j u,E,k+1)+(1-\theta_{k-1})C(S_jd,E,k+1)],\varphi_j\right\}
$$
Using Lemma \ref{lemma6-1} and Lemma \ref{lemma6-2}, our assumption of induction, and call-put symmetry (Theorem \ref{theorem6-1}), we have
\begin{eqnarray}
V_j^{k-1}&\ge&max\left\{\frac{1}{\rho_{k-1}}[\theta_{k-1}uC(S_j,E,k+1)+(1-\theta_{k-1})dC(S_j,E,k+1)],\varphi_j\right\}\nonumber
\\
&=&max\left\{\frac{1}{\eta_{k-1}}[\theta_{k-1}^\prime P(Eu,S_j,k+1)+(1-\theta_{k-1}^\prime)P(Ed,S_j,k+1)],\varphi_j\right\}\nonumber
\end{eqnarray}
From the hypotheses of the theorem and Lemma \ref{lemma3-1}, we have
$$
\rho_{k-1}\ge \rho_k,~\eta_{k-1}\le\eta_k,~\theta_{k-1}^\prime=\frac{\eta_{k-1}/\rho_{k-1}-d}{u-d}\le \theta_k^\prime
$$
and if we consider $P(Eu)\le P(Ed)$ and Lemma \ref{lemma3-2}, we obtain 
$$
\theta_{k-1}^\prime P(Eu)+(1-\theta_{k-1}^\prime)P(Ed)\ge\theta_k^\prime P(Eu)+(1-\theta_k^\prime)P(Ed).
$$
So we have
\begin{equation}\label{eq6-6}
V_j^{k-1}\ge max\left\{\frac{1}{\eta_{k-1}}[\theta_k^\prime P(Eu,S_j,k+1)+(1-\theta_k^\prime)P(Ed,S_j,k+1)],\varphi_j\right\}.
\end{equation}
Using Lemma \ref{lemma6-2}, Lemma \ref{lemma3-1} and Theorem \ref{theorem6-1} once more, we have
$$
V_j^{k-1}\ge max\left\{\frac{1}{\rho_k}[\theta_k C(S_ju,E,k+1)+(1-\theta_k)C(S_jd,E,k+1)],\varphi_j\right\}
$$
\begin{equation}\label{eq6-7}
=max\left\{\frac{1}{\rho_k}[\theta_k V_{j+1}^{k+1}+(1-\theta_k)V_{j-1}^{k+1}],\varphi_j\right\}=V_j^k.
\end{equation}
(Q.E.D)
\end{proof}

\indent

We can prove the {\bf existence of approximated optimal exercise boundary} from the monotonicity of BTM on $t$.

\begin{theorem}\label{theorem6-3}
Let $V_j^n~(n=0,1,\cdots,N,~j=0,\pm 1,\pm 2,\cdots)$ be  the price of American call option and the assumptions of Theorem \ref{theorem6-2} are satisfied. And assume that  $r(t),~q(t)>0$. Then for each $t_n (0\le n\le N-1)$, there exists $j_n$ such that 
\begin{equation}\label{eq6-8}
V_j^n=\varphi_j~for~j\ge j_n;~V_j^n>\varphi_j~for~j=j_n-1;~V_j^n\ge\varphi_j~for~j\ge j_n-2.
\end{equation}
\begin{equation}\label{eq6-9}
j_0\ge\cdots\ge j_k\ge j_{k+1}\cdots\ge j_{N-1}.
\end{equation}
\end{theorem}

{\bf Remark 6.1}
The condition $q(t)>0$ is essential and this is the difference from the condition in put option. If $q(t)=0$, then the optimal exercise boundary doesn't exist and the early exercise is not needed.
\begin{proof}
Just as Theorem \ref{theorem3-3}, we assume that $E=S_0=1,~S-j=u^j$ without loss of generality. Then
$$
V_j^N=(u^j-1)^+=\varphi_j,~(j=0,\pm 1,\pm 2,\cdots).
$$
Therefore if $j\le0$ then $\varphi_j=V_j^N=0$ and if $j\ge1$ then $\varphi_j=u^j-1=V_j^N>0$.  From
$$
V_j^{N-1}=max\left\{\frac{1}{\rho_{N-1}}[\theta_{N-1}\varphi_{j+1}+(1-\theta_{N-1})\varphi_{j-1}],\varphi_j\right\},
$$
thus if $j\le -1$ then $\varphi_{j+1}=\varphi_j=\varphi_{j-1}$ and thus $V_j^{N-1}=V_j^N=0=\varphi_j$. On the otherhand, if $j=0$ then $\varphi_1>0,~\varphi_0=\varphi_{-1}=0$ and $V_0^{N-1}=\rho_{N-1}^{-1} \varphi_1>0=\varphi_0$. So we have
\begin{equation}\label{eq6-10}
V_0^{N-1}>\varphi_0
\end{equation}
If $j\ge1$ then $j-1\ge0$ and $\varphi_j=u^j-1>0,~\varphi_{j-1}=u^{j-1}-1\ge0$,  then we have 
\begin{eqnarray}
V_j^{N-1}&=&max\left\{\frac{1}{\rho_{N-1}}[\theta_{N-1}(u^{j+1}-1)+(1-\theta_{N-1})(u^{j-1}-1)],u^j-1\right\}\nonumber
\\
&=&max\left\{\frac{1}{\rho_{N_1}}[\theta_{N-1} u^{j+1}+u^{j-1}-1-\theta_{N-1} u^{j-1}],u^j-1\right\}\nonumber
\\
&=&max\left\{\frac{1}{\rho_{N-1}}\left[ u^j \left[ u \theta_{N-1}+\frac{(1-\theta_{N-1})}{u} \right] -1 \right],u^j-1\right\}\nonumber
\\
&=&max\left\{u^j\left[\frac{u\theta_{N-1}}{\rho_{N-1}}+\frac{(1-\theta_{N-1})}{\rho_{N-1}u}\right]-\frac{1}{\rho_{N-1}},u^j-1\right\}\nonumber
\end{eqnarray}
From Lemma \ref{lemma6-1}, we have $\frac{\theta_n u}{\rho_n}+\frac{(1-\theta_n)d}{\rho_n}=\frac{1}{\eta_n}$ and thus we have
$$
V_j^{N-1}=max\left\{\frac{1}{\eta_{N-1}}u^j-\frac{1}{\rho_{N-1}},u^j-1\right\}.
$$
Since $q_{N-1}>0$, we have $\eta_{N-1}>1$ and thus if we compare graphs of $\frac{1}{\eta_{N-1}}x-\frac{1}{\rho_{N-1}}$ and $x-1$, we can see that two cases are possible.

In the case that $\forall j\ge 1,~\eta_{N-1}^{-1}u^j-\rho_{N-1}^{-1}\le u^j-1$,  we have $V_j^{N-1}=u^j-1=\varphi_j$ and thus \eqref{eq6-10} gives \eqref{eq6-8} when $j_{N-1}=1$.

On the contrary, in the case that $\exists j\ge1,~\eta_{N-1}^{-1}u^j-\rho_{N-1}^{-1}>u^j-1$, we define
$$
j_{N-1}=min\left\{j\ge1|~\eta_{N-1}^{-1}u^j-\rho_{N-1}^{-1}\le u^j-1 \right\}.
$$
Then \eqref{eq6-8} holds. That is, $j\ge j_{N-1}\Rightarrow V_j^{N-1}=\varphi_j;~j=j_{N-1}-1\Rightarrow V_j^{N-1}>\varphi_j$.  Therefore the existence of $j_{N-1}\in {\mathbf Z}$ is proved.

Assume that when $n=k$, there is $j_k(j_k\ge j_{k+1}\ge\cdots\ge j_{N-1})$ such that \eqref{eq6-8} is true, i.e $j\ge j_k\Rightarrow V_j^k=\varphi_j;~j=j_k-1\Rightarrow V_j^k>\varphi_j;~j\le j_k-2\Rightarrow V_j^k\ge\varphi_j$ (induction hypothesis). 
$$
V_j^{k-1}=max\left\{\frac{1}{\rho_{k-1}}[\theta_{k-1}V_{j+1}^k+(1-\theta_{k-1})V_{j-1}^k],\varphi_j\right\}
$$
and $j\ge j_k+1\Rightarrow j-1\ge j_k\Rightarrow V_i^k=\varphi_i,~i=j+1,j,j-1$ and so we have
\begin{eqnarray}
V_j^{k-1}&=&max\left\{\frac{1}{\rho_{k-1}}[\theta_{k-1}(u^{j+1}-1)+(1-\theta_{k-1})(u^{j-1}-1)],(u^j-1)\right\}\nonumber
\\
&=&max\left\{\frac{1}{\eta_{k-1}}u^j-\frac{1}{\rho_{k-1}},u^j-1\right\}.\nonumber
\end{eqnarray}
Since $q_{N-1}>0$, then $\eta_{k-1}>1$ and there may be the two cases, one is the case that
\begin{equation}\label{eq6-11}
\forall j\ge j_k+1,~\frac{1}{\eta_{k-1}}u^j-\frac{1}{\rho_{k-1}}\le u^j-1
\end{equation}
and the another one is the case that
\begin{equation}\label{eq6-12}
\exists j\ge j_k+1:~\frac{1}{\eta_{k-1}}u^j-\frac{1}{\rho_{k-1}}>u^j-1.
\end{equation}

If \eqref{eq6-11} is true, then $V_j^{k-1}=\varphi_j$. Thus if $j=j_k-1$, then $V_j^{k-1}\ge V_j^k>\varphi_j$. So if $V_{j_k}^{k-1}>\varphi_j$ then let $j_{k-1}=j_k+1$ and if $V_{j_k}^{k-1}=\varphi_j$ then let $j_{k-1}=j_k$.

If \eqref{eq6-12} is true, then  let 
$$
j_{k-1}=min\left\{j\ge j_k+1:~\frac{1}{\eta_{k-1}}u^j-\frac{1}{\rho_{k-1}}\le u^j-1\right\}.
$$
Then if $j\ge j_{k-1}$ then $V_j^{k-1}=\varphi_j$ and if $j=j_{k-1}-1$ then $V_j^{k-1}>\varphi_j$. If $j\le j_{k-1}-2$ then $V_j^{k-1}\ge V_j^k\ge\varphi_j$. Thus the existence of $j_{k-1}(\le j_k)$ is proved.  (Q.E.D)
\end{proof}

\indent

{\bf Remark 6.2}
Theorem \ref{theorem6-2} and Theorem \ref{theorem6-3} strongly represent the characteristics in the case of time dependent coefficients.  Especially it is remarkable that the conditions for time decreasing property are contrary to each other in put and call options.

\indent
Define the {\bf approximated optimal exercise boundary} $S=S_\Delta(t)$ on the interval $[0,T]$  as follows.

\begin{equation*}
S_\Delta(t)=\left\{
\begin{array}{ccc}
u^{j-1},~~~~~~~~~~~~~~~~~~~~~~~~~~~~~~~~~~~~~~~~~~ t=t_n
\\
\frac{(t-t_n)}{t_{n+1}-t_n}S_\Delta(t_n)+\frac{(t_{n+1}-t)}{t_{n+1}-t_n}S_\Delta(t_{
n+1}),~~t_n\le t\le t_{n+1}.
\end{array} \right.
\end{equation*}

%
%
\section{Call-put symmetry in the EDS for variational inequality model of American call option and it's applications}

\indent

The variational inequality model of American options with time dependent coefficients and the time interval partitioning method, lattice configuration and explicit difference scheme are given just as in Section 4.
In the same way as BTM, we need the homogeneity and call-put parity in order to prove the monotonicity on time of the price by the explicit difference scheme for variational inequality model of American call option.

\begin{theorem}[Homogeneity of the price by the explicit difference scheme]\label{theorem7-1}
If $u=e^{\Delta x}$, then we have
\begin{equation}\label{eq7-1}
U(\mu S_0 u^j,\mu E)=\mu U(S_0 u^j,E)~\forall \mu>0,~\forall j \in {\mathbf Z}.
\end{equation}
Here $u_j^n=u(j \Delta x+c, t_n)$ defined by \eqref{eq4-10} and \eqref{eq4-11} is denoted by $U(S_0 u^j,E;n).$
\end{theorem}

\begin{proof}
It is obvious for $n=N$ and other cases are proved by induction. (Q.E.D)
\end{proof}

\indent

Now consider the call-put parity of the price by the explicit difference scheme. Like in BTM, we can write the call and put options' prices as following.

\indent

Now consider the call-put symmetry of the price by the explicit difference scheme. Like in BTM, we can write the call and put options' prices as following.

\indent

\lefteqn{c(S,E;r,q;k)=}
\begin{equation}\label{eq7-2}
=max\left\{\frac{1}{\rho_k}[(1-\alpha)C(S,E;k+1)+\alpha(a_kC(Su,E,k+1)+(1-a_k)C(Sd,E,k+1))],(S-E)^+\right\}
\end{equation}

\lefteqn{p(S,E;r,q;k)=}
\begin{equation}\label{eq7-3}
=max\left\{\frac{1}{\rho_k}[(1-\alpha)P(S,E;k+1)+\alpha(a_kP(Su,E,k+1)+(1-a_k)P(Sd,E,k+1))],(E-S)^+\right\}
\end{equation}

{\bf Remark 7.1}
For the explicit difference scheme, the perfect symmetry  such as \eqref{eq6-3} in BTM can't be obtained.

\indent

Our goal is to prove 
\begin{equation}\label{eq7-4}
c(S,E;r,q;k)=p(E,S;q,r;k)+O(\Delta x^\delta)~~k=N,N-1,\cdots,0.
\end{equation}
Here $\delta$ will be defined later. When $k=N$, \eqref{eq7-4} holds obviously, since
$$
c(S,E;N)=(S-E)^+=p(E,S;N).
$$
In the explicit difference scheme, we use the following notations 
\begin{equation}\label{eq7-5}
a_n=\frac{1}{2}+\frac{\Delta x}{2\sigma_n^2}\left(r_n-q_n-\frac{\sigma_n^2}{2}\right),~~a_n^\prime=\frac{1}{2}+\frac{\Delta x}{2\sigma_n^2}\left(q_n-r_n-\frac{\sigma_n^2}{2}\right).
\end{equation}
Considering the homogeneity (Theorem \ref{theorem7-1}), we have
\begin{eqnarray*}
&&c(S,E;r,q,N-1)=
\\
&&=max\left\{\frac{1}{\rho_{N-1}}[(1-\alpha)c(S,E;N)+\alpha(a_{N-1}c(Su,E,N)+(1-a_{N-1})c(Sd,E;N))],(S-E)^+\right\}
\\
&&=max\left\{\frac{1}{\rho_{N-1}}c(S,E;N)+\alpha\left(\frac{a_{N-1}u}{\rho_{N-1}}c(S,Ed,N)+\frac{(1-a_{N-1})d}{\rho_{N-1}}c(S,Eu,N)\right),(S-E)^+\right\}
\\
&&=max\left\{\frac{(1-\alpha)}{\rho_{N-1}}\frac{\eta_{N-1}}{\rho_{N-1}}c(S,E;N)+\alpha\left(\frac{a_{N-1}u}{\rho_{N-1}}c(S,Ed,N)+\frac{(1-a_{N-1})d}{\rho_{N-1}}c(S,Eu,N)\right),(S-E)^+\right\}
\\
&&=max\left\{\frac{(1-\alpha)}{\eta_{N-1}}c(S,E;N)+(1-\alpha)\frac{q_{N-1}-r_{N-1}}{\rho_{N-1}}\frac{\alpha\Delta x^2}{\sigma_{N-1}^2}c(S,E;N)+\right.
\\
&&~~~~~~~~~~~\left.+\alpha\left(\frac{a_{N-1}u}{\rho_{N-1}}c(S,Ed;N)+\frac{(1-a_{N-1}d)}{\rho_{N-1}}c(S,Eu;N)\right),(S-E)^+\right\}.
\end{eqnarray*}
Now we have
\begin{eqnarray*}
\frac{a_n u}{\rho_n}-\frac{1-a_n^\prime}{\eta_n}=\frac{e^{\Delta x}}{\rho_n}\left(\frac{1}{2}+\frac{\Delta x}{2\sigma_n^2}\left(r_n-q_n-\frac{\sigma_n^2}{2}\right)\right)-\frac{1}{\eta_n}\left(\frac{1}{2}-\frac{\Delta x}{2\sigma_n^2}\left(q_n-r_n-\frac{\sigma_n^2}{2}\right)\right)
\\
=\frac{1}{\rho_n \eta_n}\left[\eta_n e^{\Delta x}\left(\frac{1}{2}+\frac{\Delta x}{2\sigma_n^2}\left(r_n-q_n-\frac{\sigma_n^2}{2}\right)\right)-\rho_n\left(\frac{1}{2}-\frac{\Delta x}{2\sigma_n^2}\left(q_n-r_n-\frac{\sigma_n^2}{2}\right)\right)\right].
\end{eqnarray*}
If we abbreviate indexes to avoid complexity, then we have
\begin{eqnarray*}
&&\eta_n e^{\Delta x}\left(\frac{1}{2}+\frac{\Delta x}{2\sigma_n^2}\left(r_n-q_n-\frac{\sigma_n^2}{2}\right)\right)=
\\
&&=\frac{1}{2}+\Delta x\left(\frac{1}{2}+\frac{1}{2\sigma_n^2}\left(r_n-q_n-\frac{\sigma_n^2}{2}\right)\right)+\Delta x^2\left(\frac{q_n \alpha}{2\sigma_n^2}+\frac{1}{4}+\frac{1}{2\sigma_n^2}\left(r_n-q_n-\frac{\sigma_n^2}{2}\right)\right)+O(\Delta x^3)
\\
&&-\rho_n\left(\frac{1}{2}-\frac{\Delta x}{2\sigma_n^2}\left(q_n-r_n-\frac{\sigma_n^2}{2}\right)\right)=\frac{1}{2}+\Delta x\left(\frac{1}{2\sigma_n^2}\left(q_n-r_n-\frac{\sigma_n^2}{2}\right)\right)-\Delta x^2\frac{r_n\alpha}{\sigma_n^2}+O(\Delta x^3).
\end{eqnarray*}
Therefore
\begin{eqnarray*}
&&\eta_n e^{\Delta x}\left(\frac{1}{2}+\frac{\Delta x}{2\sigma_n^2}\left(r_n-q_n-\frac{\sigma_n^2}{2}\right)\right)-\rho_n\left(\frac{1}{2}-\frac{\Delta x}{2\sigma_n^2}\left(q_n-r_n-\frac{\sigma_n^2}{2}\right)\right)=
\\
&&=\Delta x^2\left(\frac{q_n(\alpha-1)}{2\sigma_n^2}+\frac{r_n(1-\alpha)}{2\sigma_n^2}\right)+O(\Delta x^3)=O(\Delta x^\delta).
\end{eqnarray*}
Here
\begin{equation}\label{eq7-6}
\delta=\left\{
\begin{array}{rl}
2,~~\alpha<1
\\
3,~~\alpha=1
\end{array}
\right.
\end{equation}
Thus we have
\begin{equation}\label{eq7-7}
\frac{a_n u}{\rho_n}-\frac{1-a_n^\prime}{\eta_n}=O(\Delta x^\delta).
\end{equation}
In the same way, we have
\begin{equation}\label{eq7-8}
\frac{(1-a_n)d}{\rho_n}-\frac{a_n^\prime}{\eta_n}=\frac{\Delta x^2}{\rho_n\eta_n}\left(\frac{q_n(\alpha-1)}{2\sigma_n^2}+\frac{r_n(1-\alpha)}{2\sigma_n^2}\right)+O(\Delta x^3)=O(\Delta x^\delta).
\end{equation}
Then we obtain
\begin{eqnarray*}
&&c(S,E;r,q,N-1)=
\\
&&=max\left\{\frac{(1-\alpha)}{\eta_{N-1}}c(S,E;N)+\alpha\left(\frac{(1-a_{N-1}^\prime)}{\eta_{N-1}}c(S,Ed;N)+\frac{a_{N-1}^\prime}{\eta_{N-1}}c(S,Eu;N)\right)+O(\Delta x^\delta),(S-E)^+\right\}
\\
&&=max\left\{\frac{(1-\alpha)}{\eta_{N-1}}p(E,S;N)+\alpha\left(\frac{(1-a_{N-1}^\prime)}{\eta_{N-1}}p(Ed,S;N)+\frac{a_{N-1}^\prime}{\eta_{N-1}}p(Eu,S;N)\right),(S-E)^+\right\}+O(\Delta x^\delta)
\\
&&=max\left\{\frac{1}{\eta_{N-1}}[(1-\alpha)p(E,S;N)+\alpha(a_{N-1}^\prime p(Eu,S;N)+(1-a_{N-1}^\prime)p(Ed,S;N))],(S-E)^+\right\}+O(\Delta x^\delta)
\\
&&=p(E,S;q,r;N-1)+O(\Delta x^\delta)
\end{eqnarray*}
Therefore \eqref{eq7-4} holds true for $n=N-1$. Now inductively assume that \eqref{eq7-4} holds for $k=n+1$. From the homogeneity (Theorem \ref{theorem7-1}), we have
\begin{eqnarray*}
&&c(S,E;r,q;n)=
\\
&&=max\left\{\frac{1}{\rho_n}[(1-\alpha)c(S,E;r,q;n+1)+\alpha(a_n c(Su,E;n+1)+(1-a_n)c(Sd,E;n+1))],(S-E)^+\right\}
\\
&&=max\left\{\frac{(1-\alpha)}{\rho_n}c(S,E;r,q;n+1)+\alpha\left(\frac{a_n u}{\rho_n}c(S,Ed;n+1)+\frac{(1-a_n)d}{\rho_n}c(S,Eu;n+1)\right),(S-E)^+\right\}
\\
&&=max\left\{\frac{(1-\alpha)}{\rho_n}\frac{\eta_n}{\rho_n}c(S,E;r,q;n+1)+\alpha\left(\frac{a_n u}{\rho_n}c(S,Ed;n+1)+\frac{(1-a_n)d}{\rho_n}c(S,Eu;n+1)\right),(S-E)^+\right\}
\\
&&=max\left\{\frac{(1-\alpha)}{\eta_n}c(S,E;r,q;n+1)+(1-\alpha)\frac{q_n-r_n}{\rho_n}\frac{\alpha \Delta x^2}{\sigma_n^2}c(S,E;r,q;n+1)+\right.
\\
&&~~~~~~~\left.+\alpha\left(\frac{a_n u}{\rho_n}c(S,Ed;r,q;n+1)+\frac{(1-a_n)d}{\rho_n}c(S,Eu;r,q;n+1)\right),(S-E)^+\right\}.
\end{eqnarray*}
(Considering the equation \eqref{eq7-7} and \eqref{eq7-8})
\begin{eqnarray*}
&&=max\left\{\frac{(1-\alpha)}{\eta_n}c(S,E;r,q;n+1)+\right.
\\
&&~~~~~~\left.+\alpha\left(\frac{1-a_n^\prime}{\eta_n}c(S,Ed;r,q;n+1)+\frac{a_n^\prime}{\eta_n}c(S,Eu;r,q;n+1)\right),(S-E)^+\right\}+O(\Delta x^\delta)
\end{eqnarray*}
(Considering the induction hypothesis)
\begin{eqnarray*}
&&=max\left\{\frac{1}{\eta_n}[(1-\alpha)p(E,S;q,r;n+1)+\right.
\\
&&~~~~\left.+\alpha((1-a_n^\prime)p(Ed,S;q,r;n+1)+a_n^\prime p(Eu,S;q,r;n+1)),(S-E)^+\right\}+O(\Delta x^\delta)
\\
&&=max\left\{\frac{1}{\eta_n}[(1-\alpha)p(E,S;q,r;n+1)+\right.
\\
&&~~~~\left.+\alpha(a_n^\prime p(Eu,S;q,r;n+1)+(1-a_n^\prime)p(Ed,S;q,r;n+1)),(S-E)^+\right\}+O(\Delta x^\delta)
\\
&&=p(E,S;q,r;n)+O(\Delta x^\delta).
\end{eqnarray*}
Thus we proved the following theorem.

\begin{theorem}[Call-put parity in the explicit difference scheme]\label{theorem7-2}
$$
c(S,E;r,q;n)=p(E,S;q,r;n)+O(\Delta x^\delta).
$$
Here $\delta$ is given as \eqref{eq7-6}.
\end{theorem}

\begin{theorem}\label{theorem7-3}
Assume that $r(t)/\sigma^2(t)$ is decreasing on $t$ and $q(t)/\sigma^2(t)$ increasing on $t$. Then American call option's price $c(S,E;r,q;t_n)$ is decreasing on $t$ neglecting of $O(\Delta x^3)$, that is,
$$
c(S,E;r,q;t_n)\ge c(S,E;r,q;t_{n-1})+O(\Delta x^\delta).
$$
\end{theorem}
\begin{proof}
The decrease on $t$ is obtained by Theorem \ref{theorem4-2} and Theorem \ref{theorem7-2}.
\end{proof}
\begin{eqnarray*}
&&c(S,E;r,q;t_n)=p(E,S;q,r;t_n)+O(\Delta x^\delta)
\\
&&\ge p(E,S;q,r;t_{n+1})+O(\Delta x^\delta)=c(S,E;r,q;t_{n+1})+O(\Delta x^\delta)
\end{eqnarray*}
The first and the last equalities are from Theorem \ref{theorem7-2} and the inequality is from Theorem \ref{theorem4-2}.

\indent

{\bf Remark 7.2}
Note that unlike in the case of put options the call-put parity and $t$-decreasing property are only obtained by neglecting of infinitesimal.
Now we consider the existence of approximated optimal exercise boundary for the explicit difference scheme.

\begin{theorem}\label{theorem7-4}
Assume that $q(t)>0$, $r(t)/\sigma^2(t)$ is decreasing on $t$ and $q(t)/\sigma^2(t)$ increasing on $t$. For every $0\le n\le N-1$, there exists $j_n\in {\mathbf Z}$ such that
\begin{equation*}
j\ge j_n\Rightarrow U_j^n=\varphi_j;~j=j_n-1\Rightarrow U_j^n>\varphi_j+O(\Delta x^\delta);~ j\le j_n-2\Rightarrow U_j^n\ge\varphi_j.
\end{equation*}
\begin{equation}\label{eq7-9}
j_0\ge j_1\ge\cdots\ge j_{N-1}.
\end{equation}
\end{theorem}
\begin{proof}
First, we prove the case of $n=N-1$. $\varphi_j=(S_0 e^{j\Delta x}-E)^+$ is increasing on $j$. Let $K_1=min\{j\in{\mathbf Z}:S_0 e^{j \Delta x}-E>0\}$.  Then
\begin{equation}\label{eq7-10}
j\ge K_1\Rightarrow \varphi_j>0,~j\le K_1-1\Rightarrow\varphi_j=0.
\end{equation}
If $j\ge K_1+1$ then $j-1\ge K_1$ and $S_0 e^{j \Delta x}-E>0~(i=j+1,j,j-1)$. Let $u=e^{\Delta x},~d=e^{-\Delta x}$, then we have
\begin{eqnarray*}
U_j^{N-1}&=&max\left\{\frac{1}{\rho_{N-1}}[(1-\alpha)(S_0 e^{j \Delta x}-E)+\alpha(a_{N-1}(S_0 e^{(j+1)\Delta x}-E)+\right.
\\
&&~~~~~~~~~~\left.+(1-a_{N-1})(S_0 e^{(j-1)\Delta x}-E)],S_0 e^{j\Delta x}-E\right\}
\\
&=&max\left\{\frac{1}{\rho_{N-1}}[S_0 u^j(1-\alpha+\alpha(a_{N-1}u-(1-a_n)d))-E],S_0 u^j-E\right\}
\\
&=&max\left\{S_0 u^j\frac{(1-\alpha)+\alpha[a_{N-1}u+(1-a_{N-1}d)]}{\rho_{N-1}},S_0 u^j-E\right\}
\\
&=&max\{\psi_j,\varphi_j\}
\end{eqnarray*}
Here $\psi_j=S_0 u^j\frac{(1-\alpha)+\alpha[a_{N-1}u+(1-a_{N-1})d]}{\rho_{N-1}}-\frac{E}{\rho_{N-1}}=B S_0 u^j-\frac{E}{\rho_{N-1}}.$
$$
a_{N-1}e^{\Delta x}+(1-a_{N-1})e^{-\Delta x}=1+(r_{N-1}-q_{N-1})\frac{\Delta x^2}{\sigma_{N-1}^2}+O(\Delta x^4)
$$
and $q_{N-1}>0$, so if $\Delta x$ is small enough, then
$$
0<B=\frac{(1-\alpha)+\alpha[a_{N-1}u+(1-a_{N-1})d]}{\rho_{N-1}}=\frac{\rho_{N-1}-q_{N-1}\alpha\frac{\Delta x^2}{\sigma_{N-1}^2}+O(\Delta x^4)}{\rho_{N-1}}<1.
$$
(Here the graphs of $S_0 Bx-\frac{E}{\rho_{N-1}}$ and $S_0 x-E$ intersect.) Then we have two possibilities.

(i) The case that $\forall j\ge K_1+1,~\psi_j\le \varphi_j$. In this case $\forall j\ge K_1+1$, $U_j^{N-1}=\varphi_j$. Thus, from \eqref{eq7-10}, we have
\begin{eqnarray*}
U_{K_1-1}^{N-1}&=&max\left\{\frac{1}{\rho_{N-1}}[(1-\alpha)\varphi_{K_1-1}+\alpha(a_{N-1})\varphi_{K_1-2}],\varphi_{K_1-1}\right\}=
\\
&=&max\left\{\frac{\alpha a_{N-1}}{\rho_{N-1}}\varphi_{K_1},0\right\}>0=\varphi_{K_1-1}.
\end{eqnarray*}
From the property of American option price, we have $U_j^{N-1}\ge\varphi_j$, $j\le K_1-1$ and so if $U_{K_1}^{N-1}=\varphi_j$ then let $j_{N-1}=K_1$ and if $U_{K_1}^{N-1}>\varphi_j$ then let $j_{N-1}=K_1+1$. 

(ii) The case that $\exists j\ge K_1+1;~\psi_j>\varphi_j$. In this case $\psi_j$ is more slowly increasing than $\varphi_j$ and thus there is such integer $j_{N-1}$ that 
$$
j_{N-1}=min\{j\ge K_1+1;~\psi_j\le \varphi_j\}~(\ge K_1+2).
$$
Therefore, $j\ge j_{N-1}\Rightarrow U_j^{N-1}=\varphi_j$ and $j=j_{N-1}-1\Rightarrow\psi_j>\varphi_j$ and thus $U_j^{N-1}>\varphi_j$. From the property of American option price, $j<j_{N-1}-2\Rightarrow U_j^{N-1}\ge\varphi_j$. Thus the existence of $j_{N-1}$ in the case of $n=N-1$ is proved.

Inductively, assume that when $n=k+1$, there is such $j_{k+1}(\ge j_{k+2})$ that
$$
j\ge j_{k+1}\Rightarrow U_j^{k+1}=\varphi_j,~~j=j_{k+1}-1\Rightarrow U_j^{k+1}>\varphi_j.
$$
Recall that
$$
U_j^k=max\left\{\frac{1}{\rho_k}[(1-\alpha)U_j^{k+1}+\alpha(a_k U_{j+1}^{k+1}+(1-a_k)U_{j-1}^{k+1})],\varphi_j\right\}.
$$
Since $j\ge j_{k+1}+1\Rightarrow i=j+1,j,j-1\ge j_{k+1}\Rightarrow U_i^{k+1}=\varphi_i=S_0 e^{j\Delta x}-E$, we have
\begin{eqnarray*}
U_j^k&=&max\left\{\frac{1}{\rho_k}[(1-\alpha)\varphi_j+\alpha(a_k\varphi_{j+1}+(1-a_k)\varphi_{j-1})],\varphi_j\right\}
\\
&=&max\left\{u^j S_0\frac{1-\alpha+\alpha(a_k u+(1-a_k)d)}{\rho_k}-\frac{E}{\rho_k},S_0 u^j-E\right\}=max\{\psi_j,\varphi_j\}.
\end{eqnarray*}
Taking into account that $a_k e^{\Delta x}+(1-a_k)e^{-\Delta x}=\left(1+(r_k-q_k)\frac{\Delta x^2}{\sigma_k^2}\right)$, $q_k>0$, we have two cases just as the above.

(i) The case that $\forall j\ge j_{k+1}+1,~\psi_j\le\varphi_j$. In this case $\forall j\ge j_{k+1}+1,~U_j^k=\varphi_j$ and $j=j_{k+1}-1\Rightarrow U_j^k\ge U_j^{k+1}+O(\Delta x^\delta)>\varphi_j+O(\Delta x^\delta)$.  So if $U_{j_{k+1}}^k>\varphi_j$ then let $j_k=j_{k+1}+1$ and if $U_{j_{k+1}}^k=\varphi_j$ then let $j_k=j_{k+1}$. Then the requirement of the theorem is satisfied. 

(ii) The case that $\exists j\ge j_{k+1}+1:~\psi_j>\varphi_j$. Since $\psi_j$ is more slowly increasing than $\varphi_j$, there is such integer $j_k$ that 
$$
j_k=min\{j\ge j_{k+1}+1:~\psi_j\le \varphi_j\}~(\ge j_{k+1}+2).
$$
Then we have $j\ge j_k\Rightarrow U_j^k=\varphi_j$, $j=j_k-1\Rightarrow U_j^k>\varphi_j$. From the property of American option price $j\le j_k-2\Rightarrow U_j^k\ge \varphi_j$. Therefore the existence of $j_k$ is proved.  (Q.E.D)
\end{proof}

\indent

{\bf Remark 7.3}
Unlike in the case of put options, the existence of the optimal exercise boundary only comes from the condition $q(t)>0$ for call options. If $q(t)=0$, the optimal exercise boundary does not exist.

\indent

Now evaluate the optimal exercise boundary near maturity. In the first part in the proof of Theorem \ref{theorem7-4} we proved the existence of optimal exercise boundary $j_{N-1}$ near maturity. According to that, we have $K_1=min\{j\in {\mathbf Z}:~S_0 e^{j \Delta x}-E>0\}$ and $j\ge K_1+1\Rightarrow\varphi_j=S_0 e^{j \Delta x}-E$. Just as Theorem \ref{theorem7-4}, let 
\begin{eqnarray*}
\psi_j&=&\frac{1}{\rho_{N-1}}[(1-\alpha)\varphi_j+\alpha(a_{N-1}\varphi_{j+1}+(1-a_{N-1})\varphi_{j-1})]
\\
&=&S_0 u^j \frac{(1-\alpha)+\alpha[a_{N-1}u+(1-a_{N-1})d]}{\rho{N-1}}-\frac{E}{\rho_{N-1}}.
\end{eqnarray*}

In the case that $\forall j\ge K_1+1,~\psi_j\le \varphi_j$ we have $j_{N-1}=K_1$ or $j_{N-1}=K_1+1$ and thus $j_{N-1}\ge K_1>j_{N-1}-2$. So $S_0 e^{(j_{N-1}-2)\Delta x}\le E<S_0 e^{j_{N-1}\Delta x}$ and if we set $e^c=S_0$, then we have $j_{N-1}\Delta x+c-2\Delta x\le \ln E<j_{N-1}\Delta x+c$. So
\begin{equation}\label{eq7-11}
\ln E\le j_{N-1}\Delta x+c\le \ln E+2\Delta x.
\end{equation}
In the case that $\exists j\ge K_1=1,~\psi_j>\varphi_j$, we consider $j_{N-1}=min\{j\ge K_1+1;~\psi_j\le\varphi_j$ and 
$$
a_{N-1}e^{\Delta x}+(1-a_{N-1})e^{-\Delta x}=1+(r_{N-1}-q_{N-1})e^{-\Delta x}=1+(r_{N-1}-q_{N-1})\frac{\Delta x^2}{\sigma_{N-1}^2}+O(\Delta x^4).
$$
When $j\ge K_1+1$, we have
\begin{eqnarray*}
&&\psi_j-\varphi_j=
\\
&&=\frac{(1-\alpha)(e^{j\Delta x+c}-E)+\alpha[a_{N-1}(e^{(j+1)\Delta x+c}-E)+(1-a_{N-1})(e^{(j-1)\Delta x+c})]}{\rho_{N-1}}-(e^{j\Delta x+c}-E)
\\
&&=\frac{1}{\rho_{N-1}}\frac{\sigma_{N-1}^2\Delta t_{N-1}}{\Delta x^2}\left\{(r_{N-1}E-q_{N-1}e^{j\Delta x+c})\frac{\Delta x^2}{\sigma_{N-1}^2}+O(\Delta x^4)\right\}.
\end{eqnarray*}
and if $\Delta x$ is small enough, then we have
\begin{equation}\label{eq7-12}
\psi_j>\varphi_j\iff r_{N-1}E>q_{N-1}e^{j\Delta x+c}.
\end{equation}
On the other hand, $j\ge K_1+1\Rightarrow e^{j\Delta x+c}\ge E$ and if $q_{N-1}>r_N$ then $q_{N-1}e^{j\Delta x+c}\ge r_{N-1}E$. So $\psi_j>\varphi_j$ is not possible and in the case that $\exists j\ge K_1+1,~\psi_j>\varphi_j$, we must have $q_{N-1}\le r_{N-1}$.  From \eqref{eq7-12},
$$
j_{N-1}=min\{j\ge K_1+1:~\psi_j\le\varphi_j\}=min\{j\ge K_1+1:~r_{N-1}E\le q_{N-1}e^{j\Delta x+c}
$$
and $r_{N-1}E\le q_{N-1}e^{j_{N-1}\Delta x+c}\Rightarrow j_{N-1}\Delta x+c\ge \ln\frac{r_{N-1}}{q_{N-1}}E.$ For $j=j_{N-1}-1$ \eqref{eq7-12} holds and then $r_{N-1}E>q_{N-1}e^{(j_{N-1}-1)\Delta x+c}\Rightarrow j_{N-1}\Delta x-\Delta x+c<\ln\frac{r_{N-1}}{q_{N-1}}E.$ Thus we have
\begin{equation}\label{eq7-13}
\ln\frac{r_{N-1}}{q_{N-1}}E\le j_{N-1}\Delta x+c<\ln\frac{r_{N-1}}{q_{N-1}}E+\Delta x.
\end{equation}

By combining the inequality \eqref{eq7-11} and \eqref{eq7-13} we can obtain the following theorem, which evaluates the approximated optimal exercise boundary near maturity.

\begin{theorem}\label{theorem7-5}
$\ln max\left\{E,\frac{r_{N-1}}{q_{N-1}}E\right\}\le j_{N-1}\Delta x+c\le\ln max\left\{E,\frac{r_{N-1}}{q_{N-1}}E\right\}+2\Delta x$
\end{theorem}

{\bf Approximated optimal exercise boundary}

Fix $\Delta$ and define $\rho_{\Delta x}(t)$ as follows.
\begin{eqnarray*}
\rho_{\Delta x}=\frac{t-t_n}{t_{n+1}-t_n}(j_{n+1}\Delta x+c)+\frac{t_{n+1}-t}{t_{n+1}-t_n}(j_n\Delta x+c),
\\
~~~~~~~~~~~~~~~~~~~~~~~~~~~~t\in [t_n,t_{n+1}]~(n=0,\cdots,N-2).
\end{eqnarray*}
Then from the decreasing property of $j_n$, we know that $\rho_{\Delta x}(t)$ is decreasing.

\begin{theorem}\label{theorem7-6}

i) $\rho_{\Delta x}(t_{N-1})\in\left[\ln max\left(E,\frac{r_{N-1}}{q_{N-1}}E\right),~\ln max\left(E,\frac{r_{N-1}}{q_{N-1}}E\right)+2\Delta x\right].$

ii) $\rho_{\Delta x}(t)$ is decreasing on $t$.
\end{theorem}

%
%
\section{Convergence of the Explicit Difference Scheme and BTM}

\indent

The sequence ${\bf U}^n=(\cdots,U_j^n,\cdots)_{j=-\infty}^{\infty}$, which consists of all the prices at the points $j\Delta x+c$ of time $t_n$ given by the EDS \eqref{eq4-11}, is bounded in the sense of Theorem \ref{theorem4-1}. If we define 
\begin{equation}\label{eq8-1}
l_*^\infty({\bf Z})=\left\{{\bf U}=(U_j)_{j\in{\bf Z}}:~\left\{U_j (e^{j\Delta x+c})^{-1}\right\}:~bounded\right\},
\end{equation}
\begin{equation}\label{eq8-2}
\|{\bf Z}\|_{l_*^\infty({\bf Z})}:=\sup_j | U_j(e^{j \Delta x+c})^{-1}|,
\end{equation}
then $l_*^\infty({\bf Z})$ is a Banach space and
\begin{equation}\label{eq8-3}
{\bf U}^n=(\cdots,U_j^n,\cdots)_{j=-\infty}^\infty\in l_*^\infty({\bf Z}). 
\end{equation}
Write the right side of \eqref{eq4-11} as $({\bf F}_n{\bf U}^{n+1})_j$, i.e
\begin{equation}\label{eq8-4}
({\bf F}_n{\bf U}^{n+1})_j=max\left\{\frac{1}{\rho_n}\{(1-\alpha)U_j^{n+1}+\alpha[a_n U_{j-1}^{n+1}+(1-a_n)U_{j-1}^{n+1}]\},\varphi_j\right\},~n=N-1,\cdots,1,0.
\end{equation}
Then the operator mapping the price sequence ${\bf U}^{n+1}\in l_*^\infty({\bf Z})$ at time $t_{n+1}$ to the price sequence ${\bf U}^n=(\cdots,U_j^n,\cdots)_{j=-\infty}^\infty\in l_*^\infty({\bf Z})$ at time $t_n$ is defined:
\begin{equation}\label{eq8-5}
{\bf U}^n:={\bf F}_n {\bf U}^{n+1}=\{({\bf F}_n {\bf U}^{n+1})_j\}_{j=-\infty}^\infty 
\end{equation}
 is defined. ${\bf F}_n$ depends on $n$ and $\Delta x$ and thus it depends on $t_n,~\Delta t_n$.

\begin{lemma}\label{lemma8-1}
If $0<\alpha\le 1,~\left|\frac{\Delta x}{\sigma_n^2}\left(r_n-q_n-\frac{\sigma_n^2}{2}\right)\right|<1$, then ${\bf F}_n$ is increasing at $l_*^\infty({\bf Z})$. That is if ${\bf U}\le{\bf V}~({\bf U,V}\in l_*^\infty({\bf Z}))$,  then 
\begin{equation}\label{eq8-6}
{\bf F}_n{\bf U}\le{\bf F}_n{\bf V}.
\end{equation}
Here ${\bf U}\le{\bf V}\iff U_j\le V_j,~\forall j\in {\bf Z}$.
\end{lemma}
\begin{proof}
Under the assumption we have $(1-\alpha)\ge 0,~0<a_n<1$, and thus it can be easily proved from \eqref{eq8-4}. (Q.E.D)
\end{proof}
\begin{lemma}\label{lemma8-2}
If ${\bf U}\in l_*^\infty({\bf Z})$ and ${\bf K}=(\cdots,K,\cdots)$ is non-negative constant sequence, then 
\begin{equation}\label{eq8-7}
{\bf F}_n({\bf U}+{\bf K})\le {\bf F}_n{\bf U}+{\bf K}. 
\end{equation}
\end{lemma}
\begin{proof}
\begin{eqnarray*}
&&{\bf F}_n({\bf U}+{\bf K})=
\\
&&=\left\{max\left\{\frac{1}{\rho_n}[(1-\alpha)(U_j+K)+\alpha(a_n(U_{j+1}+K)+(1-a_n)(U_{j-1}+K))],\varphi_j\right\}\right\}_{j=-\infty}^{j=\infty}
\\
&&\le\left\{max\left\{\frac{1}{\rho_n}[(1-\alpha)U_j+\alpha(a_n U_{j+1}+(1-a_n)U_{j-1})],\varphi_j\right\}+max\left\{\frac{K}{\rho_n},0\right\}\right\}_{j=-\infty}^{j=\infty}
\\
&&\le{\bf F}_n{\bf U}+{\bf K}
\end{eqnarray*}
Here we take into account that $\rho_n>1$.  (Q.E.D)
\end{proof}

\indent

Define the approximated solution as the extension function $u_{\Delta x}(t,x)$.  When
$$
x\in \left[\left(j-\frac{1}{2}\right)\Delta x+c,\right.\left.\left(j+\frac{1}{2}\right)\Delta x+c\right),~t\in[t_n,t_{n+1})(j\in{\bf Z},~n=0,1,\cdots,N-1)
$$
We define
\begin{equation}\label{eq8-8}
u_{\Delta x}(x,t):=U_j^n. 
\end{equation}
Then from this definition and the boundedness theorem(Theorem \ref{theorem4-1}), we have
$$
0\le u_{\Delta x}(x,t)\le e^{j\Delta x+c}
$$
Thus we fix $t$ and denote
$$
{\bf u}_{\Delta x}(\bullet,~t):=\{u_{\Delta x}(x,t):~x\in{\bf R}\}\in l_*^\infty({\bf Z}).
$$
For $t\in[t_n,t_{n+1}),~n=0,\cdots,N-1$, let
\begin{equation}\label{eq8-9}
\Delta t=\frac{t-t_n}{t_{n+1}-t_n}\Delta t_{n+1}+\frac{t_{n+1}-t}{t_{n+1}-t_n}\Delta t_n. 
\end{equation}
Then $t+\Delta t\in [t_{n+1},t_{n+2})$ and from the definition \eqref{eq8-8} and \eqref{eq8-5} we have
\begin{equation}\label{eq8-10}
{\bf u}_{\Delta x}(\bullet,~t)={\bf F}_n{\bf u}_{\Delta x}(\bullet,~t+\Delta t). 
\end{equation}
\begin{theorem}[Convergence]\label{theorem8-1}
Assume that $u(x,t)$ is a viscosity solution of problem \eqref{eq4-3} for call and $r(t)/\sigma^2(t)$ is decreasing on $t$, $q(t)/\sigma^2(t)$ is increasing on$t$ and $q(t)>0$. Then we have

(i) $u_{\Delta x}(x,t)$ converges to $u(x,t)$ when $\Delta x\rightarrow 0$.

(ii) The approximated optimal exercise boundary $\rho_{\Delta x}(t)$ converges to the optimal boundary $\rho(t)$ when $\Delta x\rightarrow 0$.
\end{theorem}

 (The proof is omitted since it is similar to the proof of Theorem \ref{theorem5-2}.)
\begin{theorem}\label{theorem8-2}
{\bf (Monotonic property of American call option's price and optimal exercise boundary)}
Under the assumption of Theorem \ref{theorem8-1}, the price $V(S,t)$ of the continuous time model \eqref{eq4-1} for American call option is increasing on $S$ and decreasing on $t$. The optimal exercise boundary $\rho(t)$ is increasing on $t$.
\end{theorem}
\begin{proof}
We have the conclusions from Theorem \ref{theorem8-1}, Theorem \ref{theorem4-1}, Theorem \ref{theorem7-4}, and Theorem \ref{theorem7-6}.  (Q.E.D)
\end{proof}
\begin{theorem}[Uniformly convergence theorem]\label{theorem8-3}
Under the assumption of Theorem \ref{theorem8-1}, we have

(i) $u_{\Delta x}(x,t)$ converges to $u(x,t)$ uniformly in any compact subset of $[0,T]\times (-\infty,\infty)$.

(ii) $\rho_{\Delta x}(t)$ converges to $\rho(t)$ uniformly in $[0,T]$.
\end{theorem}
\begin{proof}
We have the conclusion from the monotonicity of American call option’s price and the optimal exercise boundary and Lemma 9 at 368p\cite{JD2}.
\end{proof}

\indent

{\bf Corollary}
 The price of BTM for American call option converges to the viscosity solution of problem \eqref{eq4-1} when $u\downarrow 1(N\rightarrow\infty)$.

 (The proof is omitted since it is similar to the proof for put option in section 5.)



\begin{thebibliography}{99}

\bibitem{ami}
Amin, K., Jump diffusion option valuation in discrete time, \textit{Journal of Finance}, \textbf{48}, 1993, 1833-1863.
\bibitem{AK}
Amin, K., Khana, A., Convergence of explicit difference scheme for American option Values from Discrete to Continuous time Financial Models, \textit{Mathematical Finance}, \textbf{4:4}(October, 1994), 289-304.
\bibitem{BDR}
Barles, G., Daher, C., Romano M., Convergence of numerical scheme for Parabolic equations arising in Finance theory, \textit{Mathematical Models and Methods in Applied Science}, \textbf{5:1}, 1995, 125-143.
\bibitem{BS}
Barles G., Souganidis, D.E., Convergence of Approximate scheme for Fully Nonlinear Second Order Equations, \textit{Asymptotic Analysis}, \textbf{1:4}, 1991, 271-283.
\bibitem{BPV}
Borovkova, S.A., Permana, F.J. and J.A.M Van Der Weide, American Basket and Spread Option Pricing by a Simple Binomial Tree, \textit{The Journal of Derivatives}, Summer 2012, 29-38.
\bibitem{CRR}
Cox, J., Ross, S., Rubinstein, M., Option Pricing: A Simplified Approach, \textit{Journal of Finanancial Economics}, \textbf{7}(October), 1979, 229-264.
\bibitem{CIL}
Crandall, M.G., Ishii, H., Lions, P.L., User’s guide to Viscosity Solutions of Second Order Partial Differential Equations, \textit{Bull. Amer. Math. Soc.}, \textbf{27}, 1992, 1-67.
\bibitem{he}
He, H., Convergence from discrete- to continuous-time contingent claims prices, \textit{Review of Financial Studies}, \textbf{3}, 1990, 523-546.
\bibitem{HL}
Hu, B., Liang, J., Optimal Convergence rate of the explicit finite difference scheme for American option Valuation, \textit{Jour. Comput. Appl.  Math.}, \textbf{230}, 2009, 583-599.
\bibitem{jia}
Jiang, L. \textit{Mathematical modeling and methods of option pricing}. Singapore: World Scientific. 2005.
\bibitem{JD1}
Jiang, L.S, Dai, M., Convergence of binomial tree methods for American Option, \textit{Proceedings of Conference on PDE and its Applications}, World Scientific, Singapore, 1999, 106-118.
\bibitem{JD2}
Jiang, L.S, Dai, M., Convergence of explicit difference scheme for American Option Valuation, \textit{Journal of Computational Mathematics}, Vol. \textbf{22}, No \textbf{3}, 2004, 371-380.
\bibitem{JD3}
Jiang, L.S, Dai, M., Convergence of binomial tree methods for European / American path dependent Option, \textit{Siam J. Numer. Anal.}, Vol. \textbf{42}, 2004, 1094-1109.
\bibitem{KOF}
Krasimir, M., Ognyan, K., F. J. Fabzzi, Y. S. KIM, and S. T. Rachev, A Binomial Tree Model for Convertible Bond Pricing, \textit{The Journal of Fixed  Income}, Winter 2013, 79-94.
\bibitem{lia}
Liang, J., On the Convergence Rate of The Binomial Tree Scheme for an American Option With Jump Diffusion, \textit{Gao Deng Xue Xiao ji Suan Shu Xue Xue Bao}, \textbf{30:1}, March 2008, 76-96. 
\bibitem{LH1}
Liang, J., Hu, B., Jiang, L.S., Bian, B.J., On the rate of Convergence of binomial tree scheme for American Option, \textit{Numer. Math.}, \textbf{107}, 2007, 333-352.
\bibitem{LH2}
Liang, J., Hu, B., Jiang, L.S., Optimal Convergence Rate of the Binomial Tree Scheme for American Options with Jump Diffusion and Their Free Boundaries, \textit{Siam J. Financial Math.}, Vol. \textbf{1}, 2010, 30-65.
\bibitem{LL}
Lin, J., Liang, J., Pricing of perpetual American and Bermudan options by binomial tree method, \textit{Front. Math. China}, \textbf{2(2)}, 2007, 243-256.
\bibitem{luo}
Luo, J. Pricing theory and application of Americal Options and Calculation of implied volatility, P.h. D Dissertation, Department of Mathematics, Tongji University, Shanghai, China, 2005.
\bibitem{O1}
O, Hyong-Chol; J.J. Jo and J.S.Kim, General Properties of Solutions to Inhomogeneous Black-Scholes Equations with Discontinuous Maturity Payoffs, Jour. Diff. Equat., \textbf{260}, Issue \textbf{4} (2016), 3151-3172. 
\bibitem{O2} 
O, Hyong-Chol; D.H. Kim; C.H.Pak, Analytical pricing of defaultable discrete coupon bonds in unified two-factor model of structural and reduced form models, \textit{Jour. Math. Anal. Appl.}, Volume \textbf{416}, Issue \textbf{1} (2014) 314 - 334.                   
\bibitem{O3} 
O, Hyong-Chol, Y.G. Kim , D.H. Kim, Higher binary with time dependent coefficients and 2 factor  model for Defaultable Bond with Discrete Default Information, \textit{Malaya Journal of Matematik}, \textbf{2(4)}, 2014, pp 330-344.
\bibitem{QXJ} 
Qian X., Xu C., Jiang L. and Bian B., Convergence of the binomial tree method for American options in jump-difusion model,~\textit{SIAM J. Numer. Anal.}, \textbf{42}, 2005, 1899-1913.
\bibitem{XQJ} 
Xu C., Qian X. and Jiang L., Numerical analysis on binomial tree methods for a jump-difusion model, \textit{Jour. Comut. Appl. Math.}, \textbf{156}, 2003, 23-45.
\bibitem{zha} 
Zhang X., Numerical analysis of American option pricing in a jump-difusion model, \textit{Math. Oper. Res.}, \textbf{22}, 1997, 668- 690.

\end{thebibliography}

\end{document}